\documentclass[runningheads]{llncs}
\usepackage{graphicx}
\usepackage{booktabs}   %% For formal tables:
\usepackage{float}
\usepackage{xspace}
\usepackage{wrapfig}
\usepackage{subcaption}
\usepackage[shortlabels]{enumitem}

\usepackage[boxed,linesnumbered]{algorithm2e}

\usepackage{cmap}

\usepackage{shuffle}
\usepackage{syntax}

\usepackage{url}
\usepackage{amsmath} % Advanced maths commands
\usepackage{amssymb} % Extra maths symbols
\usepackage{amsfonts}

\usepackage{etoolbox}
\usepackage{mathtools}
\usepackage{refcount}
\usepackage{stmaryrd}
\usepackage{enumitem}
\usepackage{listings}
\usepackage{caption}
\usepackage{flushend}

\usepackage{array}
\newcolumntype{L}[1]{>{\raggedright\let\newline\\\arraybackslash\hspace{0pt}}m{#1}}
\newcolumntype{C}[1]{>{\centering\let\newline\\\arraybackslash\hspace{0pt}}m{#1}}
\newcolumntype{R}[1]{>{\raggedleft\let\newline\\\arraybackslash\hspace{0pt}}m{#1}}

\usepackage{scalerel}
\usepackage{cancel} % for crossing arrows, etc

\usepackage{pgf}
\usepackage{tikz}
\usetikzlibrary{arrows,automata}
\usetikzlibrary{arrows.meta}
\usepackage{scalefnt}

\usepackage{wrapfig}
\usepackage{lipsum}  % generates filler text

\usepackage{algpseudocode}
\makeatletter
\renewcommand{\ALG@beginalgorithmic}{\footnotesize}
\makeatother
\usepackage{eqnarray}
\usepackage{alltt}
\usepackage{textcomp}
\usepackage{wrapfig,lipsum}
\usepackage{appendix}

\usepackage{listings}
\lstdefinelanguage{JavaScript}{%
  keywords = { async, await, break, case, catch, class, const, continue, debugger, default, delete, do, each, else, export, finally, for, function, if, import, in, instanceof, let, new, of, return, switch, this, throw, try, typeof, var, void, while, with, yield },
  morecomment = [l]{//},
  morecomment = [s]{/*}{*/},
  morestring  = [b]',
  morestring  = [b]",
  sensitive   = true,
}

\lstdefinelanguage{Java10}{
  language      = Java,
  morekeywords  ={ var },
}

\newcommand{\seqposet}{\rho}
\newcommand{\emptyseq}{\varepsilon}

\newcommand{\ops}{O}
\newcommand{\rel}{\mathcal{R}}
\newcommand{\project}[2]{{#1}_{|{#2}}}
\newcommand{\projectrv}[2]{{#1}\{{#2}\}}
\newcommand{\Meth}{\mathbb{M}}
\newcommand{\meth}{m}
\newcommand{\weaker}{\preceq}
\newcommand{\mktrueaction}[3]{{#1}({#2})\ifstrequal{#3}{\unit}{}{ \ret {#3}}}
\newcommand{\ret}{{\,\triangleright\,}}
\newcommand{\Domain}{\mathbb{D}}
\newcommand{\rv}{{\mathit{rv}}}
\newcommand{\argv}{{\mathit{arg}}}

\newcommand{\pos}{\rho}
\newcommand{\getlabel}{\ell}
\newcommand{\sequence}{sequential poset}
\newcommand{\exec}{e}
\newcommand{\causalpast}[1]{{\sf CausalHist}({#1})}
\newcommand{\causalarb}[1]{{\sf CausalArb}({#1})}
\newcommand{\causaldep}[1]{{\sf CausalPast}({#1})}
\newcommand{\poback}[1]{{\sf POPast}({#1})}
\newcommand{\axpoco}{{\sf AxCausal}}
\newcommand{\axcoarb}{{\sf AxArb}}
\newcommand{\axwcc}{{\sf AxCausalValue}}
\newcommand{\axscc}{{\sf AxCausalSeq}}
\newcommand{\axccv}{{\sf AxCausalArb}}
\newcommand{\theirarb}{\mathit{arb}}
\newcommand{\loc}{\locof{\op}}
\newcommand{\locof}[1]{\rho_{#1}}

\newcommand{\spec}{S}
\newcommand{\site}{process}
\newcommand{\sites}{processes}

\newcommand{\projrel}[2]{\project{#1}{#2}}
\newcommand{\propco}{{\sf co}}
\newcommand{\ie}{\textit{i.e.} }

\newcommand{\set}[1]{{\{{#1}\}}}

\newcommand{\tup}[1]{{\left\langle{#1}\right\rangle}}

\newcommand{\Var}{\mathsf{Var}}
\newcommand{\Val}{\mathsf{Val}}
\newcommand{\IdO}{\mathsf{IdO}}
\newcommand{\Op}{\mathsf{Op}}
\newcommand{\xvar}{{x}}
\newcommand{\yvar}{{y}}
\newcommand{\zvar}{{z}}

\newcommand{\val}{{v}}
\newcommand{\id}{{i}}
\newcommand{\rd}[3][]{\textsf{read}_{#1}({#2},{#3})}
\newcommand{\wrt}[3][]{\textsf{write}_{#1}({#2},{#3})}
\newcommand{\op}{{o}}

\newcommand{\hist}{{h}}
\newcommand{\po}{\mathsf{po}}
\newcommand{\cord}{\mathsf{co}}

\newcommand{\co}{\mathsf{co}}
\newcommand{\cf}{\mathsf{cf}}
\newcommand{\wro}[1][]{\mathsf{wr}_{#1}}

\newcommand{\hb}[1][]{\mathsf{hb}_{#1}}
\newcommand{\wop}{\mathsf{write}}
\newcommand{\rop}{\mathsf{read}}

\newcommand{\cfo}{\mathsf{cf}}

\newcommand{\readOp}[1]{\mathbb{R}({#1})}
\newcommand{\var}[1]{\mathsf{var}({#1})}

\newcommand{\writeOp}[1]{\mathbb{W}({#1})}

\renewcommand{\hist}{{h}}

\newcommand{\nClient}{\texttt{nClient}}
\newcommand{\nTransaction}{\texttt{nTransaction}}
\newcommand{\nEvent}{\texttt{nEvent}}
\newcommand{\nVariable}{\texttt{nVariable}}

\newcommand{\History}{\textsf{history}}
\newcommand{\Client}{\textsf{Client}}
\newcommand{\Transaction}{\textsf{Transaction}}
\newcommand{\Event}{\textsf{Event}}

\newcommand{\Write}{\textsf{Write}}
\newcommand{\Read}{\textsf{Read}}
\newcommand{\cm}{\texttt{CM}}
\newcommand{\cc}{\texttt{CC}}
\begin{document}
\title{Checking Causal Consistency \\of Distributed Databases\thanks{This is an extended version of a paper published in NETYS'2019 proceeding \cite{DBLP:conf/NETYS/ZennouBBEE2019}. This work is supported in part by the European Research Council (ERC) under the European Union's Horizon 2020 research and innovation programme (grant agreement No 678177). This work is also supported by Centre National pour la Recherche Scientifique et Technique (CNRST) Morocco.}}

\author{Rachid Zennou\inst{1}\inst{2} \and Ranadeep Biswas\inst{1} \and Ahmed Bouajjani\inst{1} \and \\
Constantin Enea\inst{1} \and  Mohammed Erradi\inst{2}}
\authorrunning{Rachid Zennou et al.}

\institute{Universit\'{e} de Paris, IRIF, CNRS, F-75013 Paris, France
\email{\{ranadeep,abou,cenea\}@irif.fr} 
\and
ENSIAS, Mohammed V University, Rabat, Morocco,  
\email{\{rachid.zennou,mohamed.erradi\}@gmail.com}}
%
%\maketitle              % typeset the header of the contribution

\maketitle

%!TEX root = main.tex
\begin{abstract}
%Causal consistency is one of the strongest models that can be implemented to ensure availability and partition tolerance and one of the most implemented models for distributed systems. In this paper, we propose a tool to check automatically the conformance of distributed/concurrent systems executions to causal consistency models. The idea is to reduce the problem of checking if an execution is causally consistent to Datalog queries solving using the notion of bad-patterns. Our approach was tested on several real executions of distributed databases and have shown its efficiency.
The \texttt{CAP Theorem} shows that (strong) Consistency, Availability, and Partition tolerance are impossible to be ensured together. Causal consistency is one of the %strongest
weak consistency models that can be implemented to ensure availability and partition tolerance in distributed systems. In this work, we propose a tool to check automatically the conformance of distributed/concurrent systems executions to causal consistency models. Our approach consists in reducing the problem of checking if an execution is causally consistent to solving Datalog queries. The reduction is based on complete characterizations of the executions violating causal consistency in terms of the existence of cycles in suitably defined relations between the operations occurring in these executions. We have implemented the reduction in a testing tool for distributed databases, and carried out several experiments on real case studies, showing the efficiency of the suggested approach. 

\end{abstract}

%!TEX root = main.tex
\section{Introduction}
Causal consistency \cite{DBLP:ACM/Lamport1978} is one of the most implemented models for distributed systems. Contrary to strong consistency \cite{DBLP:SIGACT/Gilbert2002} (Linearizability \cite{DBLP:ACM/Herlihy1990} and Sequential Consistency \cite{DBLP:IEEE/Lamport1979}), causal consistency can be implemented in the presence of faults while ensuring availability. Several implementations of different variants of causal consistency (such as causal convergence\cite{MISC:CAC/Mahajan2011} and causal memory \cite{DBLP:distcomp/Ahamad1995,Perrin:2016:CCB:2851141.2851170}) have been developed i.e.,\cite{DBLP:conf/SIGMOD/Bailis2013,DBLP:conf/SOCC/DU2013,DBLP:conf/SOCC/Du2014,%DBLP:journals/jss/Jimenez2008,
DBLP:sosp/Lloyd2011,DBLP:ACM/Petersen1997,DBLP:SRDSW/Preguica2014}. However, the development of such implementations that meet both consistency requirements and availability and performance requirements is an extremely hard and error prone task. Hence, developing efficient approaches to check the correctness of executions w.r.t consistency models such as causal consistency is crucial. This paper presents an approach and a tool for checking automatically the conformance of the computations of a system to causal consistency. More precisely, we address the problem of, given a computation, checking its conformance to causal consistency. We consider this problem for three variants of causal consistency that are used in practice. Solving this problem constitutes the cornerstone for developing dynamic verification and testing algorithms for causal consistency.\\
Bouajjani et al.~\cite{DBLP:conf/popl/BouajjaniEGH17} studied the
complexity of checking causal consistency for a given computation and showed that it is polynomial time. In addition, they formalized the different variations of causal consistency and proposed a reduction of this problem to the occurrence of a finite number of small "bad-patterns" in the computations (i.e., some small sets of events occurring in the computations in some particular order). In this paper, we build on that work in order to define a practical approach and a tool for checking causal consistency, and to apply this tool to real-life case studies.  Our approach consists basically in reducing the problem of detecting the existence of bad patterns defined in~\cite{DBLP:conf/popl/BouajjaniEGH17} in computations to the problem of solving a Datalog queries. The fact that solving Datalog queries is polynomial time and that our reduction is polynomial in the size of the computation, allows to solve the conformance checking for causal consistency in polynomial time ($O(n^3)$). %and match the theoretical complexity bound of the problem ($O(n^5)$).  
We have implemented our approach in an efficient testing tool for distributed systems, and carried out several experiments on real distributed databases, showing the efficiency and performance of this approach.  %To the best of our knowledge, this is the first efficient and full-automated testing tool for causal consistency verification.\\

The rest of this paper is as follows, Section \ref{sec-prel} presents preliminaries that include the used notations and the system model. Section \ref{sec-cc} is dedicated to defining the causal consistency models. Section \ref{sec-bp} recalls the characterization of causal consistency violations introduced in \cite{DBLP:conf/popl/BouajjaniEGH17}. Section \ref{sec-datalog} presents our reduction of the problem of conformance checking for causal consistency to the problem of solving Datalog queries. Section \ref{sec-res} describes our testing tool, the case studies we have considered, and the experimental results we obtained. Section \ref{sec-rw} presents related work, and finally conclusions are drown in Section \ref{sec-conc}.
%!TEX root = main.tex

\section{Preliminaries}\label{sec-prel}
\textbf{Notations.}
Given a set $\ops$ and a relation $\rel \subseteq$ $\ops$ $\times$ $\ops$, we use the notation ($o_1$,$o_2$) $\in$ $\rel$ to denote the fact that $o_1$ and $o_2$ are related by $\rel$. If $\rel$ is an order, it denotes the fact that $o_1$ precedes $o_2$ in this order. The transitive closure of $\rel$ is denoted by $\rel^+$%. The reflexive closure of $\rel$ is denoted by $\rel^*$.
, which is the composition of one or multiple copies of $\rel$

Let $\ops'$ be a subset of $\ops$. Then $\project{\rel}{\ops'}$ is
the relation $\rel$ projected on the set $\ops'$, that is
$\set{(o_1,o_2) \in \rel\ |\ o_1, o_2 \in \ops'}$.
The set $\ops' \subseteq \ops$ is said to be 
\emph{downward-closed} w.r.t a relation $\rel$ if 
$\forall o_1,o_2$, if $o_2 \in \ops'$ and $(o_1, o_2) \in \rel$ 
then $o_1 \in \ops'$ as well. 
A relation $\rel \,\, \subseteq \ops \times \ops$ is a \emph{strict partial order} if it is transitive and irreflexive. Given a strict partial order $\rel$ over $\ops$, a \emph{poset} is a pair $(\ops,\rel)$.
Notice here that we consider the strict version of posets (not the ones where the underlying partial order is \emph{weak}, \ie
reflexive, transitive and antisymmetric.
Given a set $\Sigma$, a poset $(\ops,\rel)$, and a labeling function $\getlabel: \ops \rightarrow \Sigma$, the \emph{$\Sigma$ labeled poset} $\pos$ is a tuple $(\ops,\rel,\getlabel)$.

We say that $\pos'$ is a \emph{prefix} of $\pos$ if there exists a downward 
closed set $A \subseteq O$ w.r.t. relation $\rel$ such that
$\pos' = (A,\rel,\getlabel)$. If the relation $\rel$ is a strict total order, we say that a (resp., labeled) \emph{\sequence} (sequence for short) is a (resp., labeled) poset. The concatenation of two \sequence{s} $\exec$ and $\exec'$ is denote by $\exec.\exec'$.

Consider a set of methods $\Meth$ from a domain $\Domain$.
For $\meth \in \Meth$ and $\argv,\rv \in \Domain$, and $\op \in \ops$, 
  $\getlabel(\op) = (\meth,\argv,\rv)$ means that
  operation $\op$ is an invocation of $\meth$ with input 
  $\argv$ which returns $\rv$. The label 
  $\getlabel(\op)$ is sometimes denoted $\meth(\arg,\rv)$.
  %$\mktrueaction{\meth}{\argv}{\rv}$.
Let $\rho = (\ops,\rel,\getlabel)$ be a $\Meth \times \Domain \times \Domain$ labeled poset and $\ops' \subseteq \ops$. We denote by $\projectrv{\rho}{\ops'}$ the labeled poset where we only keep the return values of the operations in $\ops'$.
Formally, $\projectrv{\rho}{\ops'}$ is the $(\Meth \times \Domain) \cup (\Meth \times \Domain \times \Domain)$ labeled poset $(\ops,\rel,\getlabel')$ where for all $\op \in \ops'$, $\getlabel'(\op) = \getlabel(\op)$, and for all $\op \in \ops \setminus \ops'$, if $\getlabel(\op) = (\meth,\argv,\rv)$, then $\getlabel'(\op) = (\meth,\argv)$. Now, we introduce a relation on labeled posets, denoted $\weaker$.
Let 
$\rho = (\ops,\rel,\getlabel)$ and 
$\rho' = (\ops,\rel',\getlabel')$ 
be two posets labeled by 
$(\Meth \times \Domain) \cup
(\Meth \times \Domain \times \Domain)$ (the return values of some operations 
in $\ops$ might not be specified). The notation $\rho' \weaker \rho$
means that $\rho'$ has less order and label constraints on the set $\ops$.
Formally, $\rho' \weaker \rho$ if $\rel'\, \subseteq\, \rel$ and for all operation 
$\op \in \ops$, 
and for all $\meth \in \Meth$, $\argv,\rv \in \Domain$,
 $\getlabel(\op) = \getlabel'(\op)$, or $\getlabel(\op) = (\meth,\argv,\rv)$ implies     $\getlabel'(\op) = (\meth,\argv)$.

\textbf{System model.}
We consider a distributed system model in which a system is composed of several processes (sites) connected over a network. Each process performs operations on objects (variables) $\Var=\{\xvar,\yvar,\ldots\}$. These objects are called replicated objects and their state is replicated at all processes. Clients interact with the system by performing operations. Assuming an unspecified set of values $\Val$ and a set of operation identifiers $\IdO$. We define the set of operations as
 $\Op=\set{\rd[\id]{\xvar}{\val},\wrt[\id]{\xvar}{\val}: \id\in\IdO, \xvar\in\Var, \val\in \Val}$. Where $\rd[\id]{\xvar}{\val}$ is a read operation reading a value $\val$ from a variable $\xvar$ and $\wrt[\id]{\xvar}{\val}$ is a write operation writing a value $\val$ on a variable $\xvar$. The set of read, resp., write, operations in a set of operations $O$ is $\readOp{O}$, resp., $\writeOp{O}$. The variable accessed by an operation $o$ is denoted by $\var{o}$.\\
\textbf{Histories.}
We consider an abstract notion of an execution called~\emph{history} which includes write and read operations. The operations performed by the same process are ordered by a \emph{program order} $\po$. We assume that histories include a \emph{write-read} relation that matches each $\textsf{read}$ operation to the $\textsf{write}$ operation written its return value. \\
Formally, a \emph{history} $\tup{O, \po, \wro}$ is a set of $\textsf{read}$ or $\textsf{write}$ operations $O$ along with a partial \emph{program order} $\po$ and a 
 \emph{write-read} relation $\wro\subseteq \writeOp{O}\times \readOp{O}$, such that if $(\wrt{\xvar}{\val},\rd{\xvar'}{\val'})\in \wro$, then $\xvar=\xvar'$ and $\val=\val'$. For $o_1$, $o_2$ $\in$ $O$, $(o_1$, $o_2)\in \po$ means that $o_1$, $o_2$  were issued by the same process and $o_1$ was submitted before $o_2$. We mention that the \emph{write-read} relation can only be defined for differentiated histories.\\
\textbf{Differentiated histories.}
A history $\tup{O, \po, \wro}$ is differentiated if each value is written at most once, i.e., for all write operations $\wrt{\xvar}{\val}$ and $\wrt{\xvar}{\val'}$, $v \ne v'$.\\
\textbf{Data Independence.}
An implementation is data-independent if its behavior does not depend on the handled values. We consider in this paper implementations that are data-independent which is a natural assumption that corresponds to a wide range of existing implementations. Under this assumption, it is good enough to consider differentiated histories~\cite{DBLP:conf/popl/BouajjaniEGH17}. Thus, all histories in this paper are differentiated. 

In addition, we assume that each history contains a write operation writing the initial value (the value 0) of variable $x$, for each variable $x$. These write operations precede all other operations in $\po$.
\\
\textbf{Specification}.
The consistency of a replicated object is defined w.r.t. some specification, determining the correct behaviors of that object in a sequential setting. In this work, we consider the read/write memory for which the specification $S_{RW}$ is inductively defined as the smallest set of sequences closed under the following rules (x $\in$ $\Var$ and v $\in$ $\Val$):
\begin{enumerate}
\item
  $\emptyseq \in S_{RW}$,
\item
  if $\seqposet \in S_{RW}$, then
  $\seqposet. \wrt{\xvar}{\val} \in S_{RW}$,

\item
  if $\seqposet \in S_{RW}$ contains no
  write on $\xvar$, then
  $\seqposet. \rd{\xvar}{0} \in S_{RW}$,
\item
   if $\seqposet \in S_{RW}$ and the last write in $\rho$ on variable $\xvar$ is
  $\wrt{\xvar}{\val}$, then
  $\seqposet. \rd{\xvar}{\val} \in S_{RW}$.
\end{enumerate}
%!TEX root = main.tex
\section{Causal Consistency}\label{sec-cc}
\subsection{Causal Consistency definitions}
Causal consistency \cite{DBLP:ACM/Lamport1978} is one of the most used models for replicated objects. It guarantees that, if two operations $\op_1$ and $\op_2$ are \emph{causally related} (some \site{} is aware of $\op_1$ when executing $\op_2$), then $\op_1$ should be executed before $\op_2$ in all \sites{}. Operations that are not causally related may be seen in different orders by different \sites{}. 
We present in the following three variations of causal consistency, weak causal consistency, causal convergence and causal memory. We use the same definitions as in \cite{DBLP:conf/popl/BouajjaniEGH17}. %Afterwards, we introduce a new formalization of the causal memory definition and the proof that it is equivalent to the one in \cite{DBLP:conf/popl/BouajjaniEGH17}.

\subsubsection{Weak causal consistency}
%%%Done
The weakest variation of causal consistency is called \emph{weak causal consistency} (\texttt{CC}, for short). A history is CC if there \emph{exists} a causal order that explains the return value of all operations in the history \cite{DBLP:conf/popl/BouajjaniEGH17}. Formally, 
\begin{definition}%\cite{DBLP:conf/popl/BouajjaniEGH17}.
A history $\hist$ satisfies CC w.r.t a specification $\spec$ if there exists a strict partial order, called  \emph{causal order}, $\co \subseteq \ops \times \ops$, such that, for all operations $\op \in \ops$ in $\hist$, there exists a specification sequence $\loc \in \spec$ such that axioms $\axpoco$ and $\axwcc$ hold (see \ref{fig:axioms}).
\end{definition}
\begin{table}

\begin{tabular}{l|l}
$\axpoco$& $\po \subseteq \co$ \\
$\axcoarb$& $\co \subseteq \theirarb$ \\
$\axwcc$& $\projectrv{\causalpast{\op}}{\op} \weaker \loc$ \\
$\axscc$& $\projectrv{\causalpast{\op}}{\poback{\op}} \weaker \loc$ \\
$\axccv$& $\projectrv{\causalarb{\op}}{\op} \weaker \loc$ \\
\end{tabular}

where:\\
$\causalpast{\op} = (\causaldep{\op},\co,\getlabel)$ \\
$\causalarb{\op} = (\causaldep{\op},\theirarb,\getlabel)$ \\
$\causaldep{\op} = \set{\op' \in \ops\ |\ (\op', \op)\in \co^*}$ \\
$\poback{\op} = \set{\op' \in \ops\ |\ (\op', \op)\in \po^*}$ \\

\caption{Axioms used in the causal consistency definitions.}
\label{fig:axioms}
\end{table}
Axiom~\axpoco{} states that the causal order should at least include the program order. 
Axiom~\axwcc{} states that, for each operation $\op \in \ops$, a valid sequence of the specification $\spec$ can be obtained by sequentializing the causal history of $\op$ i.e., all operations that precede $\op$ in the causal order. In addition, this sequentialization must also preserve the constraints provided by the causal order.

Formally, the \emph{causal past} of $\op$, $\causaldep{\op}$, is the \emph{set} of operations that precede $\op$ in the causal order. The \emph{causal history} of $\op$, $\causalpast{\op}$, is the restriction of the causal order to the operations in its causal past $\causaldep{\op}$.
The notation $\projectrv{\causalpast{\op}}{\op}$ means that only the return value of operation $\op$ is kept. The axiom~\axwcc{} uses $\projectrv{\causalpast{\op}}{\op}$ because a \site{} is not required to be consistent with the values it has returned in the past or the values returned by the other processes.

The notations $\projectrv{\causalpast{\op}}{\op} \weaker \loc$ means that $\projectrv{\causalpast{\op}}{\op}$ can be sequentialized to a sequence $\loc$ in the specification. We will formally define these last two notations in the next sections.

\setlength{\textfloatsep}{1pt}
%The weakest variation of causal consistency is called \emph{weak causal consistency} (\texttt{CC}, for short). A history is \texttt{CC} if all operations that are in a causal relation (causally-related) are seen in the same order by all processes. The relation of causality  is given by the \emph{program order} $\po$ or the \emph{write-read} relation $\wro$ or any transitive composition of these relations i.e., $\co=(\po\cup\wro)^+$. Formally, 
%\begin{definition}
%A history $\tup{O, \po, \wro}$ is \texttt{CC} if $\po\cup\wro\cup\rwo$ is acyclic.
%\end{definition}
%The \emph{read-write} relation $\rwo$ relates a read operation $\rd{\xvar}{v}$, which reads a value $v$ from a write operation $\wrt{\xvar}{v}$, with a write operation $\wrt{\xvar}{v'}$ that precedes $\wrt{\xvar}{v}$ in the causal relation. Formally, $\rwo$ is defined as
%\begin{align*}
%(\rd{\xvar}{\val},\wrt{\xvar}{\val'}) \in \rwo\mbox{ iff }&(\wrt{\xvar}{\val},\wrt{\xvar}{\val'}) \in \cord\mbox{ and } \\
%&\mbox{$(\wrt{\xvar}{\val},\rd{\xvar}{\val}) \in \wro$, for some $\wrt{\xvar}{\val}$}
%\end{align*}
For a better understanding of this model, consider the following examples.
\begin{example}
The history \ref{fig:ccnotcmnotccv} is \texttt{CC}, we can consider that $\wrt{\xvar}{1}$ is not causally-related to $\wrt{\xvar}{2}$. Therefore, $p_2$ can execute them in any order.
\end{example}
\begin{example}\label{not-cc-exe}
The history \ref{fig:notcc} is not \texttt{CC}. The reason is that, a causal order that explains the return values of all operations in the history cannot be found. Intuitively, since $\rd{\yvar}{1}$ reads the value from $\wrt{\yvar}{1}$, in any causal order, $\wrt{\yvar}{1}$ should precede $\rd{\yvar}{1}$. By transitivity of the causal order and because any causal order should include the program order, $\wrt{\xvar}{1}$ precedes $\wrt{\xvar}{2}$ in the causal order ($\wrt{\xvar}{1}$ and $\wrt{\xvar}{2}$ are causally related).
However, \site{} $p_3$ inverse this order. This is a contradiction with the informal definition of CC which requires that every \site{} should see causally related operations in the same order.

%The history \ref{fig:notcc} is not \texttt{CC}. The reason is that, we have $(\wrt{\xvar}{1}$, $\wrt{\xvar}{2}) \in \cord$ by the transitivity which include $(\wrt{\xvar}{1},\wrt{\yvar}{1}) \in \po$ and $(\wrt{\yvar}{1}$, $\rd{\yvar}{1}) \in \wro$ and $(\rd{\yvar}{1}$, $\wrt{\xvar}{2}) \in \po$. {\color{red} Then, this two writes should be seen in that order by all processes.} However, in $p_3$ we have $(\rd{\xvar}{2}, \rd{\xvar}{1}) \in \po$ which is not allowed by \texttt{CC}.
\end{example}
\begin{figure}[t]
\footnotesize
\centering
\begin{subfigure}[t]{0.4\textwidth}
\begin{minipage}[t]{0.35\textwidth}
$p_1$:\\
$\wrt{\zvar}{1}$\\
$\wrt{\xvar}{1}$\\ 
$\wrt{\yvar}{1}$\\
\end{minipage}
\begin{minipage}[t]{0.35\textwidth}
$p_2$:\\
$\wrt{\xvar}{2}$\\ 
$\rd{\zvar}{0}$\\
$\rd{\yvar}{1}$\\
$\rd{\xvar}{2}$\\
\end{minipage}
\caption{\texttt{CCv} but not \texttt{CM}}
\label{fig:ccvnotcm}
\end{subfigure}
\begin{subfigure}[t]{0.4\textwidth}
\begin{minipage}[t]{0.35\textwidth}
$p_1$:\\
$\wrt{\xvar}{1}$\\ 
$\rd{\xvar}{2}$\\
\end{minipage}
\begin{minipage}[t]{0.35\textwidth}
$p_2$:\\
$\wrt{\xvar}{2}$\\ 
$\rd{\xvar}{1}$\\
\end{minipage}
\caption{\texttt{CM} but not \texttt{CCv}}
\label{fig:cmnotccv}
\end{subfigure}
\vspace{2em}
\begin{subfigure}[t]{0.4\textwidth}
\begin{minipage}[t]{0.35\textwidth}
$p_1$:\\
$\wrt{\xvar}{1}$\\ 
$\rd{\yvar}{0}$\\
$\wrt{\yvar}{1}$\\
$\rd{\xvar}{1}$\\
\end{minipage}
\begin{minipage}[t]{0.35\textwidth}
$p_2$:\\
$\wrt{\xvar}{2}$\\ 
$\rd{\yvar}{0}$\\
$\wrt{\yvar}{2}$\\
$\rd{\xvar}{2}$\\
\end{minipage}
\caption{\texttt{CC} , \texttt{CCv} and \texttt{CM}}
\label{fig:ccccvcm}
\end{subfigure}
\begin{subfigure}[t]{0.4\textwidth}
\begin{minipage}[t]{0.35\textwidth}
$p_1$:\\
$\wrt{\xvar}{1}$\\
\end{minipage}
\begin{minipage}[t]{0.35\textwidth}
$p_2$:\\
$\wrt{\xvar}{2}$\\
$\rd{\xvar}{1}$\\
$\rd{\xvar}{2}$\\
\end{minipage}
\caption{\texttt{CC} but not \texttt{CCv} nor \texttt{CM}}
\label{fig:ccnotcmnotccv}
\end{subfigure}
\vspace{2em}
\begin{subfigure}[t]{0.6\textwidth}
\begin{minipage}[t]{0.3\textwidth}
$p_1$:\\
$\wrt{\xvar}{1}$\\ 
$\wrt{\yvar}{1}$\\
\end{minipage}
\begin{minipage}[t]{0.3\textwidth}
$p_2$:\\
$\rd{\yvar}{1}$\\
$\wrt{\xvar}{2}$\\
\end{minipage}
\begin{minipage}[t]{0.3\textwidth}
$p_3$:\\
$\rd{\xvar}{2}$\\
$\rd{\xvar}{1}$\\
\end{minipage}
\caption{not \texttt{CC} (nor \texttt{CCv}, nor \texttt{CM})}
\label{fig:notcc}
\end{subfigure}
\caption{Histories illustrating the differences between the causal consistency models \texttt{CC}, \texttt{CCv}, and \texttt{CM}.}
\label{fig:allfigs}
\end{figure} 
\subsubsection{Causal convergence}
\emph{Causal convergence} (\texttt{CCv}, for short) is stronger than \texttt{CC}. It ensures that, as long as no new updates are submitted, all \sites{} eventually converge towards the same state. In addition of seeing causally related operations in the same order (CC), causal convergence uses a total order over all the operations in a history to agree on how to order operations which are \emph{not} causally related\cite{DBLP:conf/popl/BouajjaniEGH17}. This order is called the \emph{arbitration order} and denoted by $\theirarb$. Similarly to the causal order, the arbitration order is existentially quantified in the CCv definition. Formally,
\begin{definition}%\cite{DBLP:conf/popl/BouajjaniEGH17}.
A history is \texttt{CCv} w.r.t a specification $\spec$ if there exist
a strict partial order $\co \subseteq \ops \times \ops$ and a strict total order $\theirarb \subseteq \ops \times \ops$ such that, for each operation $\op \in \ops$ in $\hist$, there exists a specification sequence $\loc \in \spec$ such that the axioms $\axpoco$, $\axcoarb$, and $\axccv$ hold.
\end{definition}
%%%%%%%%%%%%%%% 
Axiom $\axcoarb$ states that the arbitration order $\theirarb$ should at least 
include the causal order $\co$. Axiom $\axccv$ states that, sequentializing the operations that are in the causal past of $\op$ to explain the return value of an operation $\op$, should respect the arbitration order $\theirarb$.

We now present two examples, one which satisfies \texttt{CCv} and another one which violates it.
\begin{example}
The history \ref{fig:ccvnotcm} is \texttt{CCv}, we can set an arbitration order in which $\wrt{\xvar}{1}$ is ordered before $\wrt{\xvar}{2}$.
\end{example}
\begin{example}
The history \ref{fig:cmnotccv} is not \texttt{CCv}. In order to read $\rd{\xvar}{2}$, $\wrt{\xvar}{1}$ must be ordered before $\wrt{\xvar}{2}$ in the arbitration order. On the other hand, to read $\rd{\xvar}{1}$, $\wrt{\xvar}{2}$ must be ordered before $\wrt{\xvar}{1}$ in the arbitration order, that is not possible.
\end{example}
\subsubsection{Causal memory}

The third model we consider is \emph{causal memory} %~\cite{Ahamad94causalmemory,Perrin:2016:CCB:2851141.2851170} 
 (\texttt{CM}, for short) that is also stronger than \texttt{CC}. It guarantees that each process should observe concurrent operations in the same order. In addition, this order should be maintained throughout its whole execution, but it can differ from one process to another \cite{DBLP:conf/popl/BouajjaniEGH17}.
Formally, 
\begin{definition}%\cite{DBLP:conf/popl/BouajjaniEGH17}.
A history $\hist$ is CM w.r.t. a specification $\spec$ if there exists a strict partial order $\co \subseteq \ops \times \ops$ such that, for each operation $\op \in \ops$ in $\hist$, there exists a specification sequence $\loc \in \spec$ such that axioms $\axpoco$ and $\axscc$ hold.
\end{definition}

Compared to CC, CM requires that each \site{} should be consistent with the return values it has returned in the past. However, a \site{} is not required to be consistent with respect to the return values provided by other \sites{}.
Therefore, $\axscc$ states:
\[
    \projectrv{\causalpast{\op}}{\poback{\op}} \weaker \loc
\]
where $\projectrv{\causalpast{\op}}{\poback{\op}}$ is the causal history where 
we only keep the return values of the operations that precede $\op$ in the program order (in $\poback{\op}$).

As we noticed above, \texttt{CC} is weaker that \texttt{CCv} and \texttt{CM}. For instance, the history  in Figure \ref{fig:ccnotcmnotccv} is \texttt{CC} but not \texttt{CCv} nor \texttt{CM}. It is \texttt{CC}, we can consider that $\wrt{\xvar}{1}$ is not causally-related to $\wrt{\xvar}{2}$. On the other hand, for reading the value $1$ the process $p_2$ decides to order $\wrt{\xvar}{2}$ before $\wrt{\xvar}{1}$, then it changes this order to read the value $2$. This is not allowed under \texttt{CM} nor under \texttt{CCv}.\\
Both \texttt{CCv} and \texttt{CM} require that each process should observe concurrent operations in the same order. In \texttt{CM} this order can differ from one process to another. It seems that this intuitive description implies that \texttt{CCv} is stronger than \texttt{CM} but these two models are actually incomparable. The following examples illustrate the differences between these models.
\begin{example}
For instance, the history in Figure~\ref{fig:cmnotccv} is \texttt{CM}, but not \texttt{CCv}. It is not \texttt{CCv} because reading the value 1 from $x$ in the $p_1$ implies that $\wrt{\xvar}{1}$ is ordered after $\wrt{\xvar}{2}$ while reading the value 2 from $x$ in $p_2$ implies that it $\wrt{\xvar}{2}$ is ordered after $\wrt{\xvar}{1}$. This is allowed by \texttt{CM} as different processes can observe concurrent write operations in different orders.
\end{example}
\begin{example}
The history  in Figure \ref{fig:ccvnotcm} is \texttt{CCv} but not \texttt{CM}. \texttt{CCv} requires that concurrent operations should be observed in the same order by all processes. Thus, a possible order for concurrent write operations $\wrt{\xvar}{1}$ and $\wrt{\xvar}{2}$ is to order  $\wrt{\xvar}{2}$ after $\wrt{\xvar}{1}$. Under \texttt{CM}, in order to read $\rd{\zvar}{0}$, $\wrt{\xvar}{1}$ should be ordered after $\wrt{\xvar}{2}$ while to read 2 from $\xvar$, $\wrt{\xvar}{2}$ must be ordered after $\wrt{\xvar}{1}$ ($\wrt{\xvar}{1}$ must have been observed because $p_2$ reads $1$ from $\yvar$ and the writes on $x$ and $y$ are causally-related).
\end{example}

\begin{figure}[t]
% \begin{tikzpicture}
% 
\footnotesize
\centering

\includegraphics[scale=0.55]{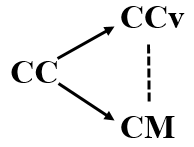} 
%\vspace{-2mm}
\caption{Relationships between causal consistency models. Directed arrows denote the ``weaker-than'' relation while dashed lines connect incomparable models.}
\label{fig:implic-cc-diagram}
%\vspace{-3mm}
\end{figure}

The Figure~\ref{fig:implic-cc-diagram} summarizes the relationships between the causal consistency models presented in this section. 

\subsection{Causal consistency violations}\label{sec-bp}
Now, we will see, for each definition of causal consistency, how to characterize histories that are not conform to causal consistency through the presence of some specific sets of operations. In \cite{DBLP:conf/popl/BouajjaniEGH17}, computations that are violations of \texttt{CC}, \texttt{CCv} or \texttt{CM} are characterized by the occurrence of a finite number of particular (small) sets of ordered events, called bad-patterns. In this section, we recall the bad-patterns corresponding to each model and their definitions (Table \ref{bp} and \ref{bpdef}). \\
\begin{table}[t]
\begin{center}
\begin{tabular}{|p{2.7cm}|p{3.3cm}|p{3.2cm}|}
\hline
   \texttt{CC} & \texttt{CCv} & \texttt{CM} \\ \hline
   \texttt{CyclicCO} & \texttt{CyclicCO} &\texttt{CyclicCO} \\
   \texttt{WriteCOInitRead} & \texttt{WriteCOInitRead} &  \texttt{WriteCOInitRead} \\
   \texttt{ThinAirRead} & \texttt{ThinAirRead} & \texttt{ThinAirRead} \\
   \texttt{WriteCORead} & \texttt{WriteCORead} & \texttt{WriteCORead}\\
   &\texttt{CyclicCF}& \texttt{WriteHBInitRead} \\
   &&\texttt{CyclicHB}\\ \hline
\end{tabular}
\vspace{.8em}
\caption{Bad-patterns for each causal consistency model}
\label{bp}
\vspace{2em}
\begin{tabular}{p{3cm} p{8cm}}
   \texttt{CyclicCO}  & the causality relation $\co$ is cylic\\
\texttt{WriteCOInitRead} & a $\rd{\xvar}{0}$ is causally preceded by a $\wrt{\xvar}{v}$ (i.e., $(\wrt{\xvar}{v}),\rd{\xvar}{0}\in \co$) such that v $\ne$ 0\\
\texttt{ThinAirRead} & there is a $\rd{\xvar}{v}$ operation that reads a value v such that v $\ne$ 0 that it is never written before i.e., it can not be related to any write by a wr relation.\\
\texttt{WriteCORead} & there exist write operations $w_1$, $w_2$ such that $\var{w_1}$ $=$ $\var{w_2}$ and a read operation $r_1$ such that $\wro(w_1,r_1)$. In addition, $(w_1,w_2) \in \co$ and $(w_2,r_1)\in \co$.\\
\texttt{WriteHBInitRead} & there exist a $\rd{\xvar}{0}$ and a $\wrt{\xvar}{v}$ (v $\ne$ 0) such that $(\wrt{\xvar}{v},\rd{\xvar}{0})\in \hb[o]$ for some operation o, with $(r,o) \in \po^*$.\\
\texttt{CyclicHB} &the $\hb[o]$ relation is cyclic for some operation o\\
\texttt{CyclicCF} &the union of $\cf$ and $\co$ (cf $\cup$ co) is cyclic \\
\end{tabular}
\end{center}
\caption{Bad-patterns definitions}
\label{bpdef}
\end{table}

%\begin{fact}
%{\color{red}
%(\cite{DBLP:conf/popl/BouajjaniEGH17})}
\subsubsection{CC Bad-patterns.}

We now give the CC bad-patterns as defined in\cite{DBLP:conf/popl/BouajjaniEGH17}. These bad-patterns are defined using the relation of causality $\co$ which is given by the \emph{program order} $\po$ or the \emph{write-read} relation $\wro$ or any transitive composition of these relations i.e., $\co=(\po\cup\wro)^+$.
\begin{lemma}(\cite{DBLP:conf/popl/BouajjaniEGH17})
A history is \texttt{CC} if it does not contain any of the bad-patterns \texttt{CyclicCO}, \texttt{WriteCOInitRead}, \texttt{ThinAirRead} and \texttt{WriteCORead}.
\end{lemma}
\begin{example}
The history in Figure \ref{fig:notcc} contains the bad-pattern\texttt{WriteCORead}. The $\wrt{\xvar}{1}$ is causally ordered before $\wrt{\xvar}{2}$ by the transitivity. On the other hand, the process $p_3$, $\rd{\xvar}{1}$ from $\wrt{\xvar}{1}$ ($\wro$($\wrt{\xvar}{1}$,$\rd{\xvar}{1}$)). The read $\rd{\xvar}{1}$ is also causally-related to $\wrt{\xvar}{2}$ by transitivity. The history in Figure \ref{fig:ccccvcm} does not contain any of the bad-patterns, so it is \texttt{CC} , \texttt{CCv} and \texttt{CM}.
\end{example}
\subsubsection{CCv bad-patterns.}
As we have seen before, CCv is stronger than CC. Therefore, CCv excludes all the CC bad-patterns we have seen above (\texttt{CyclicCO}, \texttt{WriteCOInitRead}, \texttt{ThinAirRead} and \texttt{WriteCORead}). In addition, CCv excludes another bad pattern called \texttt{CyclicCF}, defined in terms of a conflict relation $\cfo$. Intuitively, two writes $w_1$ and $w_2$ on the same variable are in conflict, if $w_1$ is causally-related to a read taking its value from $w_2$.  Formally, $\cfo$ is defined as
\begin{align*}
(\wrt{\xvar}{\val},\wrt{\xvar}{\val'}) \in \cfo \mbox{ iff }&(\wrt{\xvar}{\val},\rd{\xvar}{\val'})\in \cord\mbox{ and } \\
&\mbox{$(\wrt{\xvar}{\val'},\rd{\xvar}{\val'}) \in \wro$, for some $\rd{\xvar}{\val'}$}
\end{align*}
%\end{fact}
%\begin{fact}{\color{red}(\cite{DBLP:conf/popl/BouajjaniEGH17})}
Then,
\begin{lemma}(\cite{DBLP:conf/popl/BouajjaniEGH17})
A history is \texttt{CCv} if it is \texttt{CC} and does not contain the bad-pattern \texttt{CyclicCF}.
\end{lemma}
\begin{example}
The History in Figure \ref{fig:cmnotccv} is not \texttt{CCv} as it contains the bad-pattern \texttt{CyclicCF}. In order to read $\rd{\xvar}{2}$, $\wrt{\xvar}{2}$ must precedes $\wrt{\xvar}{2}$ in the conflict order. On the other hand, to read $\rd{\xvar}{1}$, $\wrt{\xvar}{2}$ must be ordered before $\wrt{\xvar}{1}$ in the conflict order. Thus, we get a cycle in $\cfo$.
\end{example}

\subsubsection{CM bad-Patterns.}
As we have seen above, CM is also stronger than CC. Therefore, CM excludes all the CC bad-patterns (\texttt{CyclicCO}, \texttt{WriteCOInitRead}, \texttt{ThinAirRead} and \texttt{WriteCORead}). In addition, CM excludes two additional bad-patterns (\texttt{WriteHBInitRead} and \texttt{CyclicHB}), defined using a happened-before relation per operation $\hb[o]$. Formally, $\hb[o]$ is defined as follows. 

\begin{definition}
\label{def-hb-1}
Let h=$\tup{O, \po, \wro}$ be a history. For every operation $o$ in $h$, let $\hb[o]$ be the smallest transitive relation such that:
\begin{enumerate}
    \item $\projrel{\propco}{\causaldep{\op}} \subseteq \hb{\op}$, which means that if two operations are causally related and each one is causally related to $o$,
    then they are related by $\hb[o]$ i.e., $(o_1, o_2) \in \hb[o]$ if $(o_1, o_2) \in \co$, $(o_1, o) \in \co$ and $(o_2, o) \in \co^*$ (where $\co^*$ is the reflexive closure of $\co$), and
    
    \item two writes $w_1$ and $w_2$ are related by $\hb[o]$ if $w_1$ is $\hb[o]$-related to a read taking its value from $w_2$ and that read is done by the same thread executing $o$ and before $o$, i.e., $(\wop(x, v), \wop(x, v')) \in \hb[o]$ if $(\wop(x, v), \rop(x, v')) \in \hb[o]$, $(\wop(x, v'), \rop(x, v')) \in \wro$ and $(\rop(x, v'), o) \in \po^*$ for some \linebreak $\rop(x, v')$.
\end{enumerate}
\end{definition}
%\end{fact}
%\begin{fact}{\color{red}(\cite{DBLP:conf/popl/BouajjaniEGH17})}
Then,
\begin{lemma}(\cite{DBLP:conf/popl/BouajjaniEGH17})
A history is \texttt{CM} if it is \texttt{CC} and does not contain any of the bad-patterns \texttt{WriteHBInitRead} and \texttt{CyclicHB}.
\end{lemma}
%\end{fact}

\begin{example}
\begin{enumerate}
%\item As we have seen before in the example \label{not-cc-exe}, the history in Figure \ref{fig:notcc} is not \texttt{CC}. The reason is that it contains the bad-pattern\texttt{WriteCORead}. The $\wrt{\xvar}{1}$ is causally ordered before $\wrt{\xvar}{2}$ by the transitivity. On the other hand, the process $p_3$ read $\rd{\xvar}{1}$ from $\wrt{\xvar}{1}$ ($\wro$($\wrt{\xvar}{1}$,$\rd{\xvar}{1}$)). The read $\rd{\xvar}{1}$ is also causally-related to $\wrt{\xvar}{2}$ by transitivity. 

%\item The History in Figure \ref{fig:cmnotccv} is not \texttt{CCv} as it contains the bad-pattern \texttt{CyclicCF}. In order to read $\rd{\xvar}{2}$, $\wrt{\xvar}{2}$ must precedes $\wrt{\xvar}{2}$ in the conflict order. On the other hand, to read $\rd{\xvar}{1}$, $\wrt{\xvar}{2}$ must be ordered in the conflict order before $\wrt{\xvar}{1}$. Thus lead to \texttt{CyclicCF} bad-pattern.
\item The history  \ref{fig:ccvnotcm} contains the bad-pattern \texttt{WriteHBInitRead} so it is not \texttt{CM}. Let's consider $hb=\hb[\rd{\xvar}{2}]$. We have ($\wrt{\zvar}{1}$,$\wrt{\xvar}{1}$) $\in \po$ and ($\wrt{\xvar}{1}$, $\wrt{\xvar}{2}$)  $\in hb$ (($\wrt{\xvar}{1}$,$\rd{\xvar}{2} \in \co$ implies\\ ($\wrt{\xvar}{1}$,$\wrt{\xvar}{2} \in \co$) and ($\wrt{\xvar}{2}$,$\rd{\zvar}{0}\in \po$), thus by transitivity we have ($\wrt{\zvar}{1}$,$\rd{\zvar}{0}\in hb$.
\item The history in Figure \ref{fig:ccccvcm} does not contain any of the bad-patterns, so it is \texttt{CC} , \texttt{CCv} and \texttt{CM}.
\end{enumerate}
\end{example}

\section{An improved characterization of CM}

The proposed characterization of CM in \cite{DBLP:conf/popl/BouajjaniEGH17} requires computing the $\hb[o]$ relation for all operations and then check for CM bad-patterns. Let's call this approach \texttt{CM_1}. Now, we will show that it is enough to check the CM-bad patterns for only a small set of operations (for only last operation in each process). We propose in the following a succinct but equivalent approach for checking \texttt{CM}. We show that it improves the verification runtime (Section \ref{sec-res}). Let's call this new approach \texttt{CM_2}.

To prove the equivalence between the two approaches, we have to prove some intermediate results. First, we define $\hb[o]^i$ to denote a controlled saturated version of $\hb[o]$. 

\begin{definition}
\label{def-hb-2}
Let h=$\tup{O, \po, \wro}$ be a history. For every operation $o$ in $h$,
\begin{enumerate}
    \item let $\hb[o]^0$ be the relation such that if two operations are causally related and each one is causally related to $o$, then they are related by $\hb[o]^0$ i.e., $(o_1, o_2) \in \hb[o]$ if and only if $(o_1, o_2) \in \co$, $(o_1, o) \in \co$ and $(o_2, o) \in \co^*$ (where $\co^*$ is the reflexive closure of $\co$),
    
    \item let $\hb[o]^i$ for $i > 0$ be the transitive relation if two writes $w_1$ and $w_2$ are related by $\hb[o]^i$ if $w_1$ is $(\cup_{j < i} \hb[o]^j)^+$ (transitive closure of all the previous $\hb[o]^j$) related to a read taking its value from $w_2$ and that read is done by the same thread executing $o$ and before $o$, i.e., $(\wop(x, v), \wop(x, v')) \in \hb[o]^i$ if and only if $(\wop(x, v), \rop(x, v')) \in (\cup_{j < i} \hb[o]^j)^+$, $(\wop(x, v'),\rop(x, v')) \in \wro$ and $(\rop(x, v'), o) \in \po^*$ for some $\rop(x, v')$.
\end{enumerate}
\end{definition}

\begin{theorem}
\label{thm-1}
For all $o$, $\hb[o] = (\cup_{i \geq 0} \hb[o]^i)^+$
\end{theorem}

\begin{proof}
By construction, $(\cup_{i \geq 0} \hb[o]^i)^+$ satisfies definition \ref{def-hb-1}. Because $\hb[o]$ is the smallest one, $\hb[o] \subseteq (\cup_{i \geq 0} \hb[o]^i)^+$. Also, by construction, all the relations in $(\cup_{i \geq 0} \hb[o]^i)^+$ must be present in $\hb[o]$ because they are constructed statically from $\co$ and $\wro$. So $\hb[o] \supseteq (\cup_{\geq} \hb[o]^i)^+$.
\end{proof}

Now, we prove that $\hb[o]$ is included in $\hb[o']$ if $o$ is executed before $o'$ in a same thread. Then, checking $\hb[o]$ acyclicity of $\po$-maximal operations is enough to decide for all operations. To prove this, we use the $\hb[o]^i$ definition.

\begin{lemma}
\label{lem-1}
If $(o, o') \in \po$ then $\hb[o]^i \subseteq \hb[o']^i$ for $i \geq 0$
\end{lemma}

\begin{proof}
We will prove by induction on $i$.

\begin{itemize}
    \item Base case. $i = 0$. Since $(o, o') \in \po \subseteq \co$, $(o_1, o) \in \co$ and $(o_2, o) \in \co^*$ implies, $(o_1, o') \in \co$ and $(o_2, o') \in \co^*$. So all the $\hb[o]^0 \subseteq \hb[o']^0$.
    \item Inductive step. If there exists two writes $\wop(x, v), \wop(x, v')$ and a read $\rop(x, v')$ with $(\wop(x, v), \rop(x, v')) \in (\cup_{j < i} \hb[o]^j)^+$, \\ $(\wop(x, v'),\rop(x, v')) \in \wro$  and $(\rop(x, v'), o) \in \po^*$ to force\\ $(\wop(x, v), \wop(x, v')) \in \hb[o]^i$ relation, then it is also true that,\\ $(\wop(x, v), \rop(x, v') \in (\cup_{j < i} \hb[o']^j)^+$ (induction hypothesis) and\\ $(\rop(x, v'), o') \in \po^*$. So $(\wop(x, v), \wop(x, v')) \in \hb[o']^i$ is forced as well. Thus, $\hb[o]^i \subseteq \hb[o']^i$.
\end{itemize}
\end{proof}

\begin{corollary}
\label{cor-1}
If $(o, o') \in \po$ then $\hb[o] \subseteq \hb[o']$.
\end{corollary}

\begin{proof}
Direct consequence of theorem \ref{thm-1} and lemma \ref{lem-1}.
\end{proof}

Finally, we can prove the equivalence between the two \cm{} verification approaches. Both of the approaches requires the history to be \cc{}. So, we just need to do it for the acyclicity of $\hb[o]$ and for the \texttt{WriteHBInitRead} bad-pattern.

CM_1 requires $\hb[o]$ for all $o$ to be acyclic, whereas CM_2 requires $\hb[o]$ for a subset of operations $o$ ($\po$-maximal operations) to be acyclic. So trivially CM_1 implies CM_2.

For the other direction, we use corollary \ref{cor-1}. If $(o, o') \in \po$ then $\hb[o] \subseteq \hb[o']$. Hence, a cycle in $\hb[o]$ for some $o$ (if $o$ is $\po$-maximal operation, then we are done) will be also present in $\hb[o']$ for the $\po$-maximal $o'$ after $o$ because $(o, o') \in \po$.

Now, we will prove that it is enough to check the \texttt{WriteHBInitRead} bad-pattern for only po-maximal operations as well.

Consider two operations $o_1$ and $o_2$ in a history $\hist$ such that $(o_1, o2) \in \po$. Suppose that there exists a bad-pattern \texttt{WriteHBInitRead} for $o_1$ i.e., there exist a $\rd{\xvar}{0}$ and a $\wrt{\xvar}{v}$ (v $\ne$ 0) such that $(\wrt{\xvar}{v},\rd{\xvar}{0})\in \hb[o_1]$, with $(r,o_1) \in \po^*$. Since $(o_1, o_2) \in \po$ and using the corollary \ref{cor-1}, we have $\hb[o_1] \in \hb[o_2]$, then $(\wrt{\xvar}{v},\rd{\xvar}{0})\in \hb[o_2]$ and $(r,o_2) \in \po^*$ (because of $(r,o_1) \in \po^*$). Therefore, the bad-pattern \texttt{WriteHBInitRead} exists for $o_2$ as well.

Then,
\begin{theorem}
    CM_1 and CM_2 are equivalent.
\end{theorem}

Notice that CM_2 is characterized by the same CM bad-patterns described above, it is just that the $\hb[o]$ is only computed for each po-maximal operation o in the history not for all operations (see next section).

%!TEX root = main.tex
\section{Reduction to Datalog queries solving}\label{sec-datalog}
In this section, we show our reduction of the problem of checking whether a given computation is a \texttt{CC}, \texttt{CCv} or \texttt{CM} violation to the problem of Datalog queries solving. Datalog is a logic programming language that does not allow functions as predicate arguments. The advantage of using Datalog is that it provides a high level language for naturally defining constraints on relations and that solving Datalog queries is polynomial time~\cite{MISC:STOC/Vardi1982}.
\subsection{Datalog}\label{datalog}
A rule in datalog is a statement of the following form:
\begin{center}
$r_{1}(v_{1})$ :- $r_{2}(v_{2}),...,r_{i}(v_{i}) $
\end{center}                     
Where i$\geq$ 1, $r_{i}$ are the names of predicates (relations) and $v_{i}$ are arguments. A Datalog program is a finite set of Datalog rules over the same schema \cite{MISC:foundationsofdatabases}. The LHS is called the rule head and represents the outcome of the query, while the RHS is called the rule body.
\begin{example}
For instance, this Datalog program computes the transitivity closure of a given graph. 
\begin{lstlisting}[language=Prolog]
trans(X,Y) :- edge(X,Y).
trans(X,Y)  :-  trans(X,Z), trans(Z,Y).
\end{lstlisting}
Where the fact \text{edge(a,b)} means that there exists a direct edge from a to b.
\end{example}
In the literature, there are three definitions for the semantics of Datalog programs, \textit{model theoretic}, \textit{proof-theoretic} and \textit{fixpoint semantics} \cite{MISC:foundationsofdatabases}. In this paper, we have considered the \textit{fix-point semantics}.
\subsubsection{Fix-point semantics.}
This approach is based on the fix-point theory. Given a function $f()$, its fix-point is an element $e$ from its domain which is mapped by the function $f$ to itself i.e., $f(e)=e$. Each Datalog program has an associated operator called immediate consequence operator. Applying repeatedly this operator on existing facts generates new facts until getting a fixed point. 
\subsection{Histories Encoding}
In our approach, extracted relations from a history  ($\po$, $\wro$...) are represented as predicates called facts, while the algorithm for fixed point computation is formulated as Datalog recursive relations called inference rules.\\
We first introduce all the facts. For instance, consider the fact $\po$(a,b) which represents the program order from the operation a to the operation b (similarly  $\po$(b,c)),
\begin{lstlisting}[language=Prolog]
po(a,b).
po(b,c).
\end{lstlisting}
Now, we define the necessary relations (axioms) for our approach.
\begin{itemize}
\item \textit{rd(X)}, X is a read operation
\item \textit{wrt(X)}, X is a write operation
\item \textit{po(X,Y)}, X precedes Y in the $\po$ order.
\item \textit{wr(X,Y)}, Y reads the value from a write operation X ($\wro$ relation)
%\item \textit{co(X,Y)}, X precedes Y in the causal order.
\item \textit{sv(X,Y)}, the operations X and Y accessed to the same variable.
\end{itemize} 
Then, we define the inference rules used to generate derived relations. For instance, the following rule states that the causal relation $\co$ is transitive.
\begin{lstlisting}[language=Prolog]
co(X,Z) :- co(X,Y), co(Y,Z).
\end{lstlisting}
\subsection{Bad-patterns Encoding}
We have expressed all the bad-patterns as Datalog inference rules, except\\ \texttt{ThinAirRead} that we verify externally as it contains a universal quantification over all operations. There exist two kinds of bad-patterns. The first type is related to the existence of a cycle in a relation. For instance, the bad-pattern \texttt{CyclicCO} that can be expressed as
\begin{lstlisting}[language=Prolog]
:- co(X,Y), co(Y,X).
\end{lstlisting}
Intuitively, this means that there exist no operations X and Y such that X precedes Y in the causal order and Y also precedes X in the causal order.
Since $\cord$ is transitive, we can simply write it as
\begin{lstlisting}[language=Prolog]
:- co(X,X).
\end{lstlisting}
The second type is related to the occurrence of some operations in some particular order. For instance, \texttt{WriteCORead} can be expressed as follows
\begin{lstlisting}[language=Prolog]
:- co(X,Y), co(Y,Z), wr(X,Z), wrt(X), wrt(Y), rd(Z), sv(X,Y), sv(Y,Z). 
\end{lstlisting}
Intuitively, this means that there exist no write operations X and Y on the same variable and a read operation Z which takes the value from X such that X precedes Y in the causal order and Y precedes Z in the causal order.\\
\subsubsection{CC bad-patterns encoding.}
In addition of the \texttt{CyclicCO} bad-pattern we have seen above, we will see how the other CC bad-patterns are encoded. Consider the following example which presents the Datalog program corresponding to an execution.
\begin{example}\label{hisexdg}
This example represents the history \ref{fig:cmnotccv} Datalog program for checking CC: 
\begin{lstlisting}[language=Prolog]
% Facts
wrt("w(x,1,id0)").
po("w(x,1,id0)","r(x,2,id1)").
sv("r(x,2,id1)","w(x,1,id0)").
sv("w(x,2,id2)","w(x,1,id0)").
sv("r(x,1,id3)","w(x,1,id0)").
rd("r(x,2,id1)").
sv("w(x,1,id0)","r(x,2,id1)").
wr("w(x,2,id2)","r(x,2,id1)").
sv("w(x,2,id2)","r(x,2,id1)").
sv("r(x,1,id3)","r(x,2,id1)").
wrt("w(x,2,id2)").
sv("w(x,1,id0)","w(x,2,id2)").
sv("r(x,2,id1)","w(x,2,id2)").
po("w(x,2,id2)","r(x,1,id3)").
sv("r(x,1,id3)","w(x,2,id2)").
rd("r(x,1,id3)").
wr("w(x,1,id0)","r(x,1,id3)").
sv("w(x,1,id0)","r(x,1,id3)").
sv("r(x,2,id1)","r(x,1,id3)").
sv("w(x,2,id2)","r(x,1,id3)").
initread("r(a,0,ida)"). 
% Inference rules
co(X,Y) :- po(X,Y).
co(X,Y) :- wr(X,Y).
co(X,Z) :- co(X,Y), co(Y,Z). % Transitivity
% CC bad-patterns
:- co(X,X). % CyclicCO
:- co(X,Y), wrt(X), initread(Y), sv(X,Y). % WriteCOInitRead
:- co(X,Y), co(Y,Z), wr(X,Z), wrt(X), wrt(Y), rd(Z), sv(X,Y), sv(Y,Z).
% WriteCORead
\end{lstlisting}
\end{example}
Notice that, since the bad pattern WriteCOInitRead includes a predicate initread(Y), we add the initread("r(a,0,ida)") to the programs that do not contain a read which reads the initial value.

The result of running this Datalog program using the online clingo version (https://potassco.org/clingo/run/) is shown in the following:

\begin{lstlisting}[language=Prolog]
clingo version 5.5.0
Reading from stdin
Solving...
Answer: 1
po("w(x,1,id0)","r(x,2,id1)") po("w(x,2,id2)","r(x,1,id3)") co("w(x,1,id0)","r(x,2,id1)") co("w(x,2,id2)","r(x,1,id3)") co("w(x,2,id2)","r(x,2,id1)") co("w(x,1,id0)","r(x,1,id3)") wr("w(x,2,id2)","r(x,2,id1)") wr("w(x,1,id0)","r(x,1,id3)") sv("r(x,2,id1)","w(x,1,id0)") sv("w(x,2,id2)","w(x,1,id0)") sv("r(x,1,id3)","w(x,1,id0)") sv("w(x,1,id0)","r(x,2,id1)") sv("w(x,2,id2)","r(x,2,id1)") sv("r(x,1,id3)","r(x,2,id1)") sv("w(x,1,id0)","w(x,2,id2)") sv("r(x,2,id1)","w(x,2,id2)") sv("r(x,1,id3)","w(x,2,id2)") sv("w(x,1,id0)","r(x,1,id3)") sv("r(x,2,id1)","r(x,1,id3)") sv("w(x,2,id2)","r(x,1,id3)") initread("r(a,0,ida)") wrt("w(x,1,id0)") wrt("w(x,2,id2)") rd("r(x,2,id1)") rd("r(x,1,id3)")
SATISFIABLE

Models       : 1
Calls        : 1
Time         : 0.008s (Solving: 0.00s 1st Model: 0.00s Unsat: 0.00s)
CPU Time     : 0.000s
\end{lstlisting}

Now, let's see how CCv and CM bad-patterns are encoded. Since the CCv/CM bad-patterns include CC bad-patterns, each CCv/CM Datalog program should contain CC bad-patterns which we have already seen above in addition of some other rules we will see in the next sections.
\subsubsection{CCv bad-patterns encoding.}
The CCv bad-patterns are encoded as follows:
\begin{lstlisting}[language=Prolog]
% CCv bad-patterns
cf(X,Y) :- co(X,Z), wr(Y,Z), wrt(X), sv(X,Y), sv(X,Z). %Conflict order CF
cf(X,Y) :- cf(X,Z), cf(Z,Y). %Transitivity
cfco(X,Y) :- co(X,Y). %cfco= CF U CO, cfco is the union of cf and co.
cfco(X,Y) :- cf(X,Y). %cfco= CF U CO 
%CCv bad-pattern
 :- cfco(X,Y), cfco(Y,X). %CyclicCF CF U CO
 \end{lstlisting}

Let's consider an example of CCv Datalog programs. As we have seen, the example \ref{fig:cmnotccv} is not CCv so the following Datalog program for checking CCv is not satisfiable.

\begin{lstlisting}[language=Prolog]
% Facts
wrt("w(x,1,id0)").
po("w(x,1,id0)","r(x,2,id1)").
sv("r(x,2,id1)","w(x,1,id0)").
sv("w(x,2,id2)","w(x,1,id0)").
sv("r(x,1,id3)","w(x,1,id0)").
rd("r(x,2,id1)").
sv("w(x,1,id0)","r(x,2,id1)").
wr("w(x,2,id2)","r(x,2,id1)").
sv("w(x,2,id2)","r(x,2,id1)").
sv("r(x,1,id3)","r(x,2,id1)").
wrt("w(x,2,id2)").
sv("w(x,1,id0)","w(x,2,id2)").
sv("r(x,2,id1)","w(x,2,id2)").
po("w(x,2,id2)","r(x,1,id3)").
sv("r(x,1,id3)","w(x,2,id2)").
rd("r(x,1,id3)").
wr("w(x,1,id0)","r(x,1,id3)").
sv("w(x,1,id0)","r(x,1,id3)").
sv("r(x,2,id1)","r(x,1,id3)").
sv("w(x,2,id2)","r(x,1,id3)").
initread("r(a,0,ida)"). 
% CC inference rules
co(X,Y) :- po(X,Y).
co(X,Y) :- wr(X,Y).
co(X,Z) :- co(X,Y), co(Y,Z). % Transitivity
% CC bad-patterns
:- co(X,X). % CyclicCO
:- co(X,Y), wrt(X), initread(Y), sv(X,Y). % WriteCOInitRead
:- co(X,Y), co(Y,Z), wr(X,Z), wrt(X), wrt(Y), rd(Z), sv(X,Y), sv(Y,Z).
% WriteCORead

% CCv inference rules
cf(X,Y) :- co(X,Z), wr(Y,Z), wrt(X), sv(X,Y), sv(X,Z). %Conflict order CF
cf(X,Y) :- cf(X,Z), cf(Z,Y). %Transitivity
cfco(X,Y) :- co(X,Y). %cfco= CF U CO, cfco is the union of cf and co.
cfco(X,Y) :- cf(X,Y). %cfco= CF U CO
%CCv bad-pattern
 :- cfco(X,Y), cfco(Y,X). %CyclicCF CF U CO
 
 clingo version 5.5.0
Reading from stdin
Solving...
UNSATISFIABLE

Models       : 0
Calls        : 1
Time         : 0.009s 
CPU Time     : 0.000s
\end{lstlisting}

\subsubsection{CM bad-patterns encoding.}
The CM bad-patterns are encoded as follows:

\begin{lstlisting}[language=Prolog]

%CM inference rules
hb(X,O,O) :- co(X,O). %hbo initialized to causal order
hb(X,Y,O) :- hb(Y,O,O), co(X,Y). 
hb(X,Y,O) :- hb(X,Z,O), po(Z,O), wr(Y,Z), wrt(X), sv(X,Y).% hbo definition
hb(X,Y,O) :- hb(X,Z,O), wr(Y,Z), wrt(X), sv(X,Y). % hbo definition
hb(X,Z,O) :- hb(X,Y,O), hb(Y,Z,O). %Transitivity
%CM bad-patterns
:- hb(X,Y,O), wrt(X), sv(X,Y), po(Y,O), initread(Y). %WriteHBInitRead
:- hb(X,Y,O), hb(Y,X,O). %CyclicHB
\end{lstlisting}

As we have mentioned in the section 5, CM_1 and CM_2 are characterized by the same CM bad-patterns described above. The only difference is that for the CM_2, we have added a function which identifies the po-maximal operation in each thread. We replace then the operation "$O$" in the CM bad-patterns above by these identified operations (last operation in each thread) instead of replacing it by all read/write operations in the history (CM_1).

For a better understanding, consider the instantiation of the CM bad-patterns for CM_1 and CM_2.

\begin{itemize}
\item For CM_1: We replace "$O$" by all operations in the history.
\end{itemize}
\begin{lstlisting}[language=Prolog]

%CM inference rules

hb(X,w(x,1,id0),w(x,1,id0)) :- co(X,w(x,1,id0)). 
hb(X,Y,w(x,1,id0)) :- hb(Y,w(x,1,id0),w(x,1,id0)), co(X,Y). 
hb(X,Y,w(x,1,id0)) :- hb(X,Z,w(x,1,id0)), po(Z,w(x,1,id0)), wr(Y,Z), wrt(X), sv(X,Y).
hb(X,Y,w(x,1,id0)) :- hb(X,Z,w(x,1,id0)), wr(Y,Z), wrt(X), sv(X,Y). 
hb(X,Z,w(x,1,id0)) :- hb(X,Y,w(x,1,id0)), hb(Y,Z,w(x,1,id0)). 
%CM bad-patterns
:- hb(X,Y,w(x,1,id0)), wrt(X), sv(X,Y), po(Y,w(x,1,id0)), initread(Y). 
:- hb(X,Y,w(x,1,id0)), hb(Y,X,w(x,1,id0)). 
%CM inference rules
hb(X,w(x,2,id2),w(x,2,id2)) :- co(X,w(x,2,id2)). 
hb(X,Y,w(x,2,id2)) :- hb(Y,w(x,2,id2),w(x,2,id2)), co(X,Y). 
hb(X,Y,w(x,2,id2)) :- hb(X,Z,w(x,2,id2)), po(Z,w(x,2,id2)), wr(Y,Z), wrt(X), sv(X,Y).
hb(X,Y,w(x,2,id2)) :- hb(X,Z,w(x,2,id2)), wr(Y,Z), wrt(X), sv(X,Y). 
hb(X,Z,w(x,2,id2)) :- hb(X,Y,w(x,2,id2)), hb(Y,Z,w(x,2,id2)). 
%CM bad-patterns
:- hb(X,Y,w(x,2,id2)), wrt(X), sv(X,Y), po(Y,w(x,2,id2)), initread(Y). 
:- hb(X,Y,w(x,2,id2)), hb(Y,X,w(x,2,id2)). 

%CM inference rules
hb(X,"r(x,2,id1)","r(x,2,id1)") :- co(X,"r(x,2,id1)").
hb(X,Y,"r(x,2,id1)") :- hb(Y,"r(x,2,id1)","r(x,2,id1)"), co(X,Y). 
hb(X,Y,"r(x,2,id1)") :- hb(X,Z,"r(x,2,id1)"), po(Z,"r(x,2,id1)"), wr(Y,Z), wrt(X), sv(X,Y).
hb(X,Y,"r(x,2,id1)") :- hb(X,Z,"r(x,2,id1)"), wr(Y,Z), wrt(X), sv(X,Y).
hb(X,Z,"r(x,2,id1)") :- hb(X,Y,"r(x,2,id1)"), hb(Y,Z,"r(x,2,id1)").
%CM bad-patterns
:- hb(X,Y,"r(x,2,id1)"), wrt(X), sv(X,Y), po(Y,"r(x,2,id1)"), initread(Y). 
:- hb(X,Y,"r(x,2,id1)"), hb(Y,X,"r(x,2,id1)"). 
%CM inference rules
hb(X,"r(x,1,id3)","r(x,1,id3)") :- co(X,"r(x,1,id3)"). 
hb(X,Y,"r(x,1,id3)") :- hb(Y,"r(x,1,id3)","r(x,1,id3)"), co(X,Y). 
hb(X,Y,"r(x,1,id3)") :- hb(X,Z,"r(x,1,id3)"), po(Z,"r(x,1,id3)"), wr(Y,Z), wrt(X), sv(X,Y).
hb(X,Y,"r(x,1,id3)") :- hb(X,Z,"r(x,1,id3)"), wr(Y,Z), wrt(X), sv(X,Y). 
hb(X,Z,"r(x,1,id3)") :- hb(X,Y,"r(x,1,id3)"), hb(Y,Z,"r(x,1,id3)"). 
%CM bad-patterns
:- hb(X,Y,"r(x,1,id3)"), wrt(X), sv(X,Y), po(Y,"r(x,1,id3)"), initread(Y). 
:- hb(X,Y,"r(x,1,id3)"), hb(Y,X,"r(x,1,id3)"). 
\end{lstlisting}

\begin{itemize}
\item For CM_2: We replace "$O$" by the last operation in each process in the history.
\end{itemize}
\begin{lstlisting}[language=Prolog]
%CM inference rules
hb(X,"r(x,2,id1)","r(x,2,id1)") :- co(X,"r(x,2,id1)").
hb(X,Y,"r(x,2,id1)") :- hb(Y,"r(x,2,id1)","r(x,2,id1)"), co(X,Y). 
hb(X,Y,"r(x,2,id1)") :- hb(X,Z,"r(x,2,id1)"), po(Z,"r(x,2,id1)"), wr(Y,Z), wrt(X), sv(X,Y).
hb(X,Y,"r(x,2,id1)") :- hb(X,Z,"r(x,2,id1)"), wr(Y,Z), wrt(X), sv(X,Y).
hb(X,Z,"r(x,2,id1)") :- hb(X,Y,"r(x,2,id1)"), hb(Y,Z,"r(x,2,id1)").
%CM bad-patterns
:- hb(X,Y,"r(x,2,id1)"), wrt(X), sv(X,Y), po(Y,"r(x,2,id1)"), initread(Y). 
:- hb(X,Y,"r(x,2,id1)"), hb(Y,X,"r(x,2,id1)"). 
%CM inference rules
hb(X,"r(x,1,id3)","r(x,1,id3)") :- co(X,"r(x,1,id3)"). 
hb(X,Y,"r(x,1,id3)") :- hb(Y,"r(x,1,id3)","r(x,1,id3)"), co(X,Y). 
hb(X,Y,"r(x,1,id3)") :- hb(X,Z,"r(x,1,id3)"), po(Z,"r(x,1,id3)"), wr(Y,Z), wrt(X), sv(X,Y).
hb(X,Y,"r(x,1,id3)") :- hb(X,Z,"r(x,1,id3)"), wr(Y,Z), wrt(X), sv(X,Y). 
hb(X,Z,"r(x,1,id3)") :- hb(X,Y,"r(x,1,id3)"), hb(Y,Z,"r(x,1,id3)"). 
%CM bad-patterns
:- hb(X,Y,"r(x,1,id3)"), wrt(X), sv(X,Y), po(Y,"r(x,1,id3)"), initread(Y). 
:- hb(X,Y,"r(x,1,id3)"), hb(Y,X,"r(x,1,id3)"). 
\end{lstlisting}
%\end{itemize}

Now let's consider the whole Datalog programs and their running results.
\begin{itemize}
\item For CM_1:
\end{itemize}
\begin{lstlisting}[language=Prolog]
% Facts
wrt("w(x,1,id0)").
po("w(x,1,id0)","r(x,2,id1)").
sv("r(x,2,id1)","w(x,1,id0)").
sv("w(x,2,id2)","w(x,1,id0)").
sv("r(x,1,id3)","w(x,1,id0)").
rd("r(x,2,id1)").
sv("w(x,1,id0)","r(x,2,id1)").
wr("w(x,2,id2)","r(x,2,id1)").
sv("w(x,2,id2)","r(x,2,id1)").
sv("r(x,1,id3)","r(x,2,id1)").
wrt("w(x,2,id2)").
sv("w(x,1,id0)","w(x,2,id2)").
sv("r(x,2,id1)","w(x,2,id2)").
po("w(x,2,id2)","r(x,1,id3)").
sv("r(x,1,id3)","w(x,2,id2)").
rd("r(x,1,id3)").
wr("w(x,1,id0)","r(x,1,id3)").
sv("w(x,1,id0)","r(x,1,id3)").
sv("r(x,2,id1)","r(x,1,id3)").
sv("w(x,2,id2)","r(x,1,id3)").
initread("r(a,0,ida)"). 
% Inference rules
co(X,Y) :- po(X,Y).
co(X,Y) :- wr(X,Y).
co(X,Z) :- co(X,Y), co(Y,Z). % Transitivity
% CC bad-patterns
:- co(X,X). % CyclicCO
:- co(X,Y), wrt(X), initread(Y), sv(X,Y). % WriteCOInitRead
:- co(X,Y), co(Y,Z), wr(X,Z), wrt(X), wrt(Y), rd(Z), sv(X,Y), sv(Y,Z).
% WriteCORead


%CM inference rules
hb(X,w(x,1,id0),w(x,1,id0)) :- co(X,w(x,1,id0)). 
hb(X,Y,w(x,1,id0)) :- hb(Y,w(x,1,id0),w(x,1,id0)), co(X,Y). 
hb(X,Y,w(x,1,id0)) :- hb(X,Z,w(x,1,id0)), po(Z,w(x,1,id0)), wr(Y,Z), wrt(X), sv(X,Y).
hb(X,Y,w(x,1,id0)) :- hb(X,Z,w(x,1,id0)), wr(Y,Z), wrt(X), sv(X,Y). 
hb(X,Z,w(x,1,id0)) :- hb(X,Y,w(x,1,id0)), hb(Y,Z,w(x,1,id0)). 
%CM bad-patterns
:- hb(X,Y,w(x,1,id0)), wrt(X), sv(X,Y), po(Y,w(x,1,id0)), initread(Y). 
:- hb(X,Y,w(x,1,id0)), hb(Y,X,w(x,1,id0)). 
% CM inference rules
hb(X,w(x,2,id2),w(x,2,id2)) :- co(X,w(x,2,id2)). 
hb(X,Y,w(x,2,id2)) :- hb(Y,w(x,2,id2),w(x,2,id2)), co(X,Y). 
hb(X,Y,w(x,2,id2)) :- hb(X,Z,w(x,2,id2)), po(Z,w(x,2,id2)), wr(Y,Z), wrt(X), sv(X,Y).
hb(X,Y,w(x,2,id2)) :- hb(X,Z,w(x,2,id2)), wr(Y,Z), wrt(X), sv(X,Y). 
hb(X,Z,w(x,2,id2)) :- hb(X,Y,w(x,2,id2)), hb(Y,Z,w(x,2,id2)). 
%CM bad-patterns
:- hb(X,Y,w(x,2,id2)), wrt(X), sv(X,Y), po(Y,w(x,2,id2)), initread(Y). 
:- hb(X,Y,w(x,2,id2)), hb(Y,X,w(x,2,id2)). 
% CM inference rules
hb(X,"r(x,2,id1)","r(x,2,id1)") :- co(X,"r(x,2,id1)").
hb(X,Y,"r(x,2,id1)") :- hb(Y,"r(x,2,id1)","r(x,2,id1)"), co(X,Y). 
hb(X,Y,"r(x,2,id1)") :- hb(X,Z,"r(x,2,id1)"), po(Z,"r(x,2,id1)"), wr(Y,Z), wrt(X), sv(X,Y).
hb(X,Y,"r(x,2,id1)") :- hb(X,Z,"r(x,2,id1)"), wr(Y,Z), wrt(X), sv(X,Y).
hb(X,Z,"r(x,2,id1)") :- hb(X,Y,"r(x,2,id1)"), hb(Y,Z,"r(x,2,id1)").
%CM bad-patterns
:- hb(X,Y,"r(x,2,id1)"), wrt(X), sv(X,Y), po(Y,"r(x,2,id1)"), initread(Y). 
:- hb(X,Y,"r(x,2,id1)"), hb(Y,X,"r(x,2,id1)"). 
% CM inference rules
hb(X,"r(x,1,id3)","r(x,1,id3)") :- co(X,"r(x,1,id3)"). 
hb(X,Y,"r(x,1,id3)") :- hb(Y,"r(x,1,id3)","r(x,1,id3)"), co(X,Y). 
hb(X,Y,"r(x,1,id3)") :- hb(X,Z,"r(x,1,id3)"), po(Z,"r(x,1,id3)"), wr(Y,Z), wrt(X), sv(X,Y).
hb(X,Y,"r(x,1,id3)") :- hb(X,Z,"r(x,1,id3)"), wr(Y,Z), wrt(X), sv(X,Y). 
hb(X,Z,"r(x,1,id3)") :- hb(X,Y,"r(x,1,id3)"), hb(Y,Z,"r(x,1,id3)"). 
%CM bad-patterns
:- hb(X,Y,"r(x,1,id3)"), wrt(X), sv(X,Y), po(Y,"r(x,1,id3)"), initread(Y). 
:- hb(X,Y,"r(x,1,id3)"), hb(Y,X,"r(x,1,id3)"). 

clingo version 5.5.0
Reading from stdin
Solving...
Answer: 1
po("w(x,1,id0)","r(x,2,id1)") po("w(x,2,id2)","r(x,1,id3)") co("w(x,1,id0)","r(x,2,id1)") co("w(x,2,id2)","r(x,1,id3)") co("w(x,2,id2)","r(x,2,id1)") co("w(x,1,id0)","r(x,1,id3)") wr("w(x,2,id2)","r(x,2,id1)") wr("w(x,1,id0)","r(x,1,id3)") sv("r(x,2,id1)","w(x,1,id0)") sv("w(x,2,id2)","w(x,1,id0)") sv("r(x,1,id3)","w(x,1,id0)") sv("w(x,1,id0)","r(x,2,id1)") sv("w(x,2,id2)","r(x,2,id1)") sv("r(x,1,id3)","r(x,2,id1)") sv("w(x,1,id0)","w(x,2,id2)") sv("r(x,2,id1)","w(x,2,id2)") sv("r(x,1,id3)","w(x,2,id2)") sv("w(x,1,id0)","r(x,1,id3)") sv("r(x,2,id1)","r(x,1,id3)") sv("w(x,2,id2)","r(x,1,id3)") initread("r(a,0,ida)") wrt("w(x,1,id0)") wrt("w(x,2,id2)") rd("r(x,2,id1)") rd("r(x,1,id3)") hb("w(x,2,id2)","r(x,1,id3)","r(x,1,id3)") hb("w(x,1,id0)","r(x,1,id3)","r(x,1,id3)") hb("w(x,2,id2)","w(x,1,id0)","r(x,1,id3)") hb("w(x,1,id0)","r(x,2,id1)","r(x,2,id1)") hb("w(x,2,id2)","r(x,2,id1)","r(x,2,id1)") hb("w(x,1,id0)","w(x,2,id2)","r(x,2,id1)")
SATISFIABLE

Models       : 1
Calls        : 1
Time         : 0.029s
CPU Time     : 0.000s
\end{lstlisting}

\begin{itemize}
\item For CM_2:
\end{itemize}
\begin{lstlisting}[language=Prolog]
% Facts
wrt("w(x,1,id0)").
po("w(x,1,id0)","r(x,2,id1)").
sv("r(x,2,id1)","w(x,1,id0)").
sv("w(x,2,id2)","w(x,1,id0)").
sv("r(x,1,id3)","w(x,1,id0)").
rd("r(x,2,id1)").
sv("w(x,1,id0)","r(x,2,id1)").
wr("w(x,2,id2)","r(x,2,id1)").
sv("w(x,2,id2)","r(x,2,id1)").
sv("r(x,1,id3)","r(x,2,id1)").
wrt("w(x,2,id2)").
sv("w(x,1,id0)","w(x,2,id2)").
sv("r(x,2,id1)","w(x,2,id2)").
po("w(x,2,id2)","r(x,1,id3)").
sv("r(x,1,id3)","w(x,2,id2)").
rd("r(x,1,id3)").
wr("w(x,1,id0)","r(x,1,id3)").
sv("w(x,1,id0)","r(x,1,id3)").
sv("r(x,2,id1)","r(x,1,id3)").
sv("w(x,2,id2)","r(x,1,id3)").
initread("r(a,0,ida)"). 
% CC inference rules
co(X,Y) :- po(X,Y).
co(X,Y) :- wr(X,Y).
co(X,Z) :- co(X,Y), co(Y,Z). % Transitivity
% CC bad-patterns
:- co(X,X). % CyclicCO
:- co(X,Y), wrt(X), initread(Y), sv(X,Y). % WriteCOInitRead
:- co(X,Y), co(Y,Z), wr(X,Z), wrt(X), wrt(Y), rd(Z), sv(X,Y), sv(Y,Z).
% WriteCORead

% CM inference rules

hb(X,"r(x,2,id1)","r(x,2,id1)") :- co(X,"r(x,2,id1)").
hb(X,Y,"r(x,2,id1)") :- hb(Y,"r(x,2,id1)","r(x,2,id1)"), co(X,Y). 
hb(X,Y,"r(x,2,id1)") :- hb(X,Z,"r(x,2,id1)"), po(Z,"r(x,2,id1)"), wr(Y,Z), wrt(X), sv(X,Y).
hb(X,Y,"r(x,2,id1)") :- hb(X,Z,"r(x,2,id1)"), wr(Y,Z), wrt(X), sv(X,Y).
hb(X,Z,"r(x,2,id1)") :- hb(X,Y,"r(x,2,id1)"), hb(Y,Z,"r(x,2,id1)").
%CM bad-patterns
:- hb(X,Y,"r(x,2,id1)"), wrt(X), sv(X,Y), po(Y,"r(x,2,id1)"), initread(Y). 
:- hb(X,Y,"r(x,2,id1)"), hb(Y,X,"r(x,2,id1)").
 
% CM inference rules
hb(X,"r(x,1,id3)","r(x,1,id3)") :- co(X,"r(x,1,id3)"). 
hb(X,Y,"r(x,1,id3)") :- hb(Y,"r(x,1,id3)","r(x,1,id3)"), co(X,Y). 
hb(X,Y,"r(x,1,id3)") :- hb(X,Z,"r(x,1,id3)"), po(Z,"r(x,1,id3)"), wr(Y,Z), wrt(X), sv(X,Y).
hb(X,Y,"r(x,1,id3)") :- hb(X,Z,"r(x,1,id3)"), wr(Y,Z), wrt(X), sv(X,Y). 
hb(X,Z,"r(x,1,id3)") :- hb(X,Y,"r(x,1,id3)"), hb(Y,Z,"r(x,1,id3)"). 
%CM bad-patterns
:- hb(X,Y,"r(x,1,id3)"), wrt(X), sv(X,Y), po(Y,"r(x,1,id3)"), initread(Y). 
:- hb(X,Y,"r(x,1,id3)"), hb(Y,X,"r(x,1,id3)").

clingo version 5.5.0
Reading from stdin
Solving...
Answer: 1
po("w(x,1,id0)","r(x,2,id1)") po("w(x,2,id2)","r(x,1,id3)") co("w(x,1,id0)","r(x,2,id1)") co("w(x,2,id2)","r(x,1,id3)") co("w(x,2,id2)","r(x,2,id1)") co("w(x,1,id0)","r(x,1,id3)") wr("w(x,2,id2)","r(x,2,id1)") wr("w(x,1,id0)","r(x,1,id3)") sv("r(x,2,id1)","w(x,1,id0)") sv("w(x,2,id2)","w(x,1,id0)") sv("r(x,1,id3)","w(x,1,id0)") sv("w(x,1,id0)","r(x,2,id1)") sv("w(x,2,id2)","r(x,2,id1)") sv("r(x,1,id3)","r(x,2,id1)") sv("w(x,1,id0)","w(x,2,id2)") sv("r(x,2,id1)","w(x,2,id2)") sv("r(x,1,id3)","w(x,2,id2)") sv("w(x,1,id0)","r(x,1,id3)") sv("r(x,2,id1)","r(x,1,id3)") sv("w(x,2,id2)","r(x,1,id3)") initread("r(a,0,ida)") wrt("w(x,1,id0)") wrt("w(x,2,id2)") rd("r(x,2,id1)") rd("r(x,1,id3)") hb("w(x,2,id2)","r(x,1,id3)","r(x,1,id3)") hb("w(x,1,id0)","r(x,1,id3)","r(x,1,id3)") hb("w(x,2,id2)","w(x,1,id0)","r(x,1,id3)") hb("w(x,1,id0)","r(x,2,id1)","r(x,2,id1)") hb("w(x,2,id2)","r(x,2,id1)","r(x,2,id1)") hb("w(x,1,id0)","w(x,2,id2)","r(x,2,id1)")
SATISFIABLE

Models       : 1
Calls        : 1
Time         : 0.011s 
CPU Time     : 0.000s
\end{lstlisting}

As we have mentioned before, our new approach (CM_2) computes the  $\hb[o]$ relation for a small set of operations (po-maximal operations) compared to CM_1. As can be seen above, the size of the Datalog program was considerably reduced when we use CM_2 for a small history. %Now, one can imagine the effect of this on 
Let alone long histories that contains hundreds of operations. This will be seen in the experimental results (Section \ref{sec-res}).
\subsection{A procedure for checking Causal Consistency}
Let's name the procedure which implements the reduction we have seen in the previous section REDUC-to-DATALOG. This procedure takes as input the history $\hist$ and the causal consistency model $\mathcal{M}$ to check, and returns the corresponding Datalog program $\mathcal{D}$. Then, we call another procedure named DATALOG-SOLVER which verifies whether $\mathcal{D}$ is SATISFIABLE or not.

\begin{algorithm}
{\footnotesize
 \SetKwInOut{KwInput}{Input}
 \SetKwInOut{KwOutput}{Output}
 \KwIn{A history $\hist = \tup{O, \po, \wro}$ and a causal consistency model \texttt{M}} 
 \KwOut{\texttt{SAT} iff $\hist$ satisfies \texttt{M}}
 \BlankLine
 REDUC-to-DATALOG(h,M)\\
 \eIf{ DATALOG-SOLVER(REDUC-to-DATALOG(h,M))}{
   \Return{\texttt{true}}\;
   }{
    \Return{\texttt{false}}\;
  }
 }
 \caption{Checking Causal Consistency.}
 \label{ccalgo}
\end{algorithm}
\begin{theorem}
Algorithm 1 returns \texttt{true} iff the input history $\hist$ satisfies the causal consistency model M.
\end{theorem}
The correctness of this theorem is ensured by the fact that our reduction is a simple and direct encoding of bad patterns to Datalog and these bad-patterns were proven in~\cite{DBLP:conf/popl/BouajjaniEGH17} to capture exactly the causal consistency violations.
 \subsection{Complexity}
 The complexity of a Datalog program is $\mathcal{O}(n^k)$~\cite{DBLP:DATALOGCOMP/W99}, where n is the number of constants in the input data, and k is the maximum number of variables in a clause. As we have seen in the previous section, given a history $\tup{O, \po, \wro}$, the maximum number of variables in a rule in our Datalog programs is 3, thus the complexity of our approach is $\mathcal{O}(n^3)$, where n is the size of the computation (the number of operations). Our approach's complexity is better than the one defined in~\cite{DBLP:conf/popl/BouajjaniEGH17} in which the complexity of checking CC, CCv and CM was shown to be $\mathcal{O}(n^5)$. 
%Our reduction is polynomial time in the size of the computation. For a given execution, the relations po and wr can be extracted directly (as all the considered execution are differentiated) and their size is relative to the computation size. Moreover, the size of bad-patterns is constant on the execution and the complexity of evaluating a Datalog programs is PTIME~\cite{MISC:STOC/Vardi1982}. Thus, the complexity of our approach is PTIME, which meets the complexity shown in~\cite{DBLP:conf/popl/BouajjaniEGH17}.
%!TEX root = main.tex
\section{Experimental Evaluation}\label{sec-res}
\setlength{\textfloatsep}{1pt}
\setlength{\intextsep}{7pt}
We have investigated the efficiency and scalability of our tool (named 	\textit{CausalC-Checker}) by applying it to two real-life distributed transactional databases, CockroachDB \cite{COCKRDB} and Galera \cite{GALDB}.
\begin{figure}[hb]
\centering
\includegraphics[width=1\textwidth]{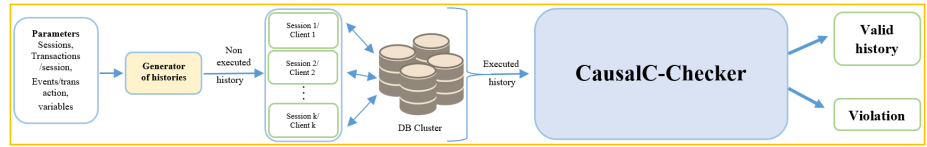}
\caption{The General architecture of the histories checking procedure}
\label{fig:arcgen}
\end{figure}
\begin{algorithm}[!htb]
 \SetKwInOut{KwInput}{Input}
 \SetKwInOut{KwOutput}{Output}
 \KwIn{$\nClient$, $\nTransaction$, $\nEvent$, $\nVariable$}
 \KwOut{A non executed $\History$}
 \BlankLine
 $lastWrite \leftarrow \emptyset$\;
 \ForEach{$v \in 1..\nVariable$}{
  $lastWrite(v) \leftarrow 0$\;
 }
 $\History \leftarrow \emptyset$\;
 \ForEach{$1..\nClient$}{
  $\Client \leftarrow \emptyset$\;
  \ForEach{$1..\nTransaction$}{
   $\Transaction \leftarrow \emptyset$\;
   \ForEach{$1..\nEvent$}{
    $\Event \leftarrow new(\Event)$\;
    $\Event.operation \leftarrow uniformly\_choose(\{\Read, \Write\})$\;
    $\Event.variable \leftarrow uniformly\_choose(\{1..\nVariable\})$\;
    \If{$\Event.operation = \Write$}{
     $\Event.value \leftarrow lastWrite(\Event.variable) + 1$\;
     $lastWrite(\Event.variable) \leftarrow lastWrite(\Event.variable) + 1$\;
    }
    $\Transaction.push(\Event)$\;
   }
   $\Client.push(\Transaction)$\;
  }
  $\History.push(\Client)$\;
 }
 \Return{$\History$}\;

\caption{The histories generator algorithm}
\label{hisgen}
\end{algorithm}

\textbf{Histories generation:} The Figure \ref{fig:arcgen} presents the general architecture of the used testing procedure in the next experiments.
Histories are generated using random clients with the parameters, the number of sessions, the number of transactions per session, the number of events per transaction (in this paper, we consider one event per transaction), and the number of variables. A client is generated by the generator of histories (Algorithm \ref{hisgen}) by choosing randomly the type of operation (read or write) in each transaction, the variable and a value for write operations. That constitutes non executed histories that are the histories which do not contain the return values of read operations. Each client performs a session, communicates with the database cluster by executing operations (read/write) and gets the return values for read operations. The recorded histories are called executed histories in the Figure \ref{fig:arcgen}. We ensure that all histories are differentiated. These histories are the input of our \textit{CausalC-Checker}. %In the following, the \texttt{CM} definition in \cite{DBLP:conf/popl/BouajjaniEGH17} (definition \ref{def-cm-1}) is called \texttt{CM_1} and the new definition (definition \ref{def-cm-2}) is named \texttt{CM_2}.

\subsection{Case study 1: CockroachDB.}
We have used the highly available and strongly consistent distributed database CockroachDB \cite{COCKRDB} (v2.1.0) that is built on a transactional strongly-consistent key-value store, so it is expected to be causally consistent. Considering one operation per transaction lead to our model.\\
\begin{figure*}
\footnotesize
\centering
\begin{subfigure}[t]{1\textwidth}
\begin{minipage}[t]{0.45\textwidth}
\includegraphics[scale=0.235]{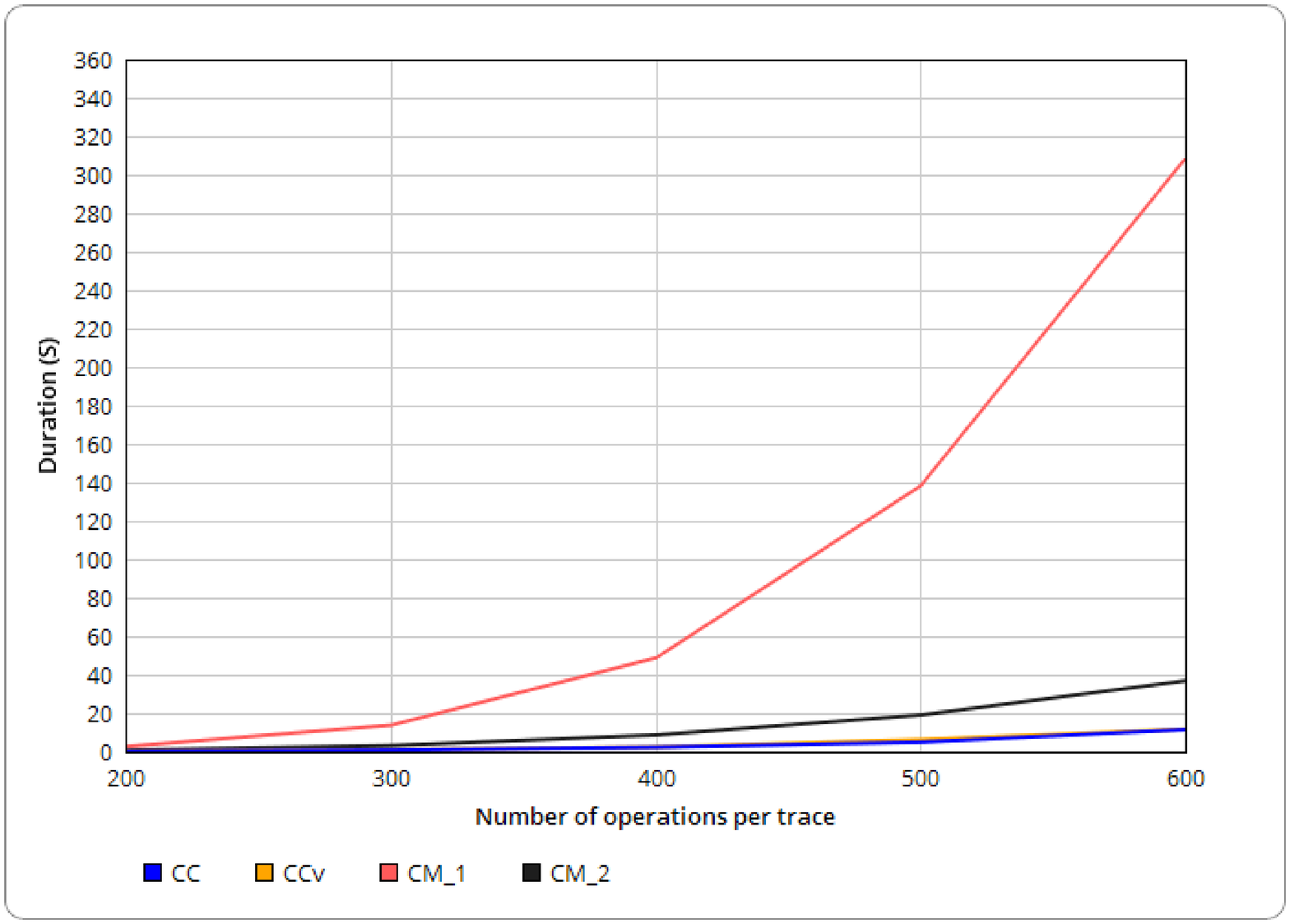} 
\caption{Checking Causal Consistency while varying the number of operations.}
\label{fig:ccccvcm-ops}
\end{minipage}
\hspace{3mm}
\begin{minipage}[t]{0.45\textwidth}
\includegraphics[scale=0.235]{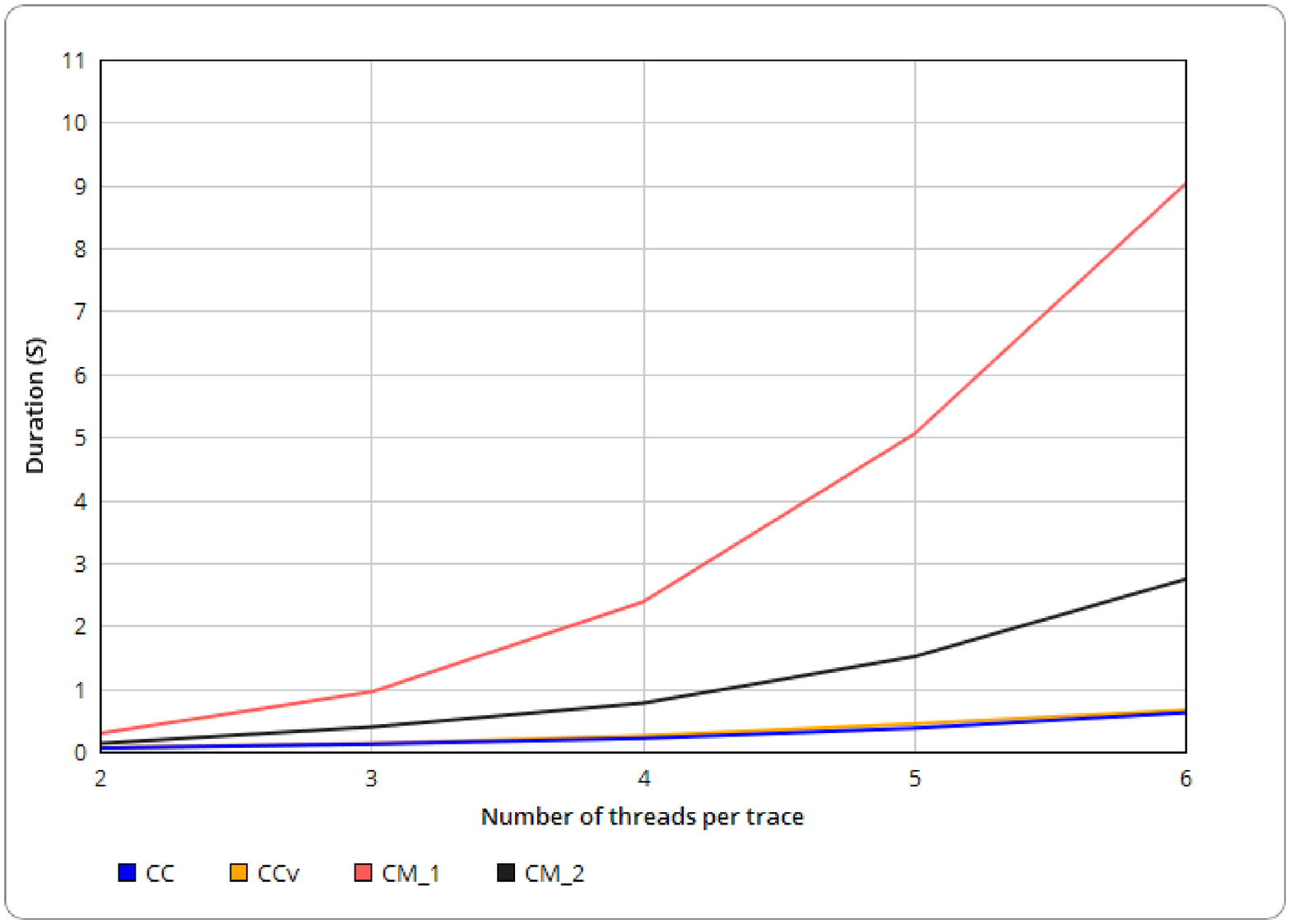} 
\caption{Checking Causal Consistency while varying the number of processes.}
\label{fig:ccccvcm-cpus}
\end{minipage}
\hspace{3mm}
\end{subfigure}
\begin{subfigure}[t]{1\textwidth}
\begin{minipage}[t]{0.45\textwidth}
\includegraphics[scale=0.235]{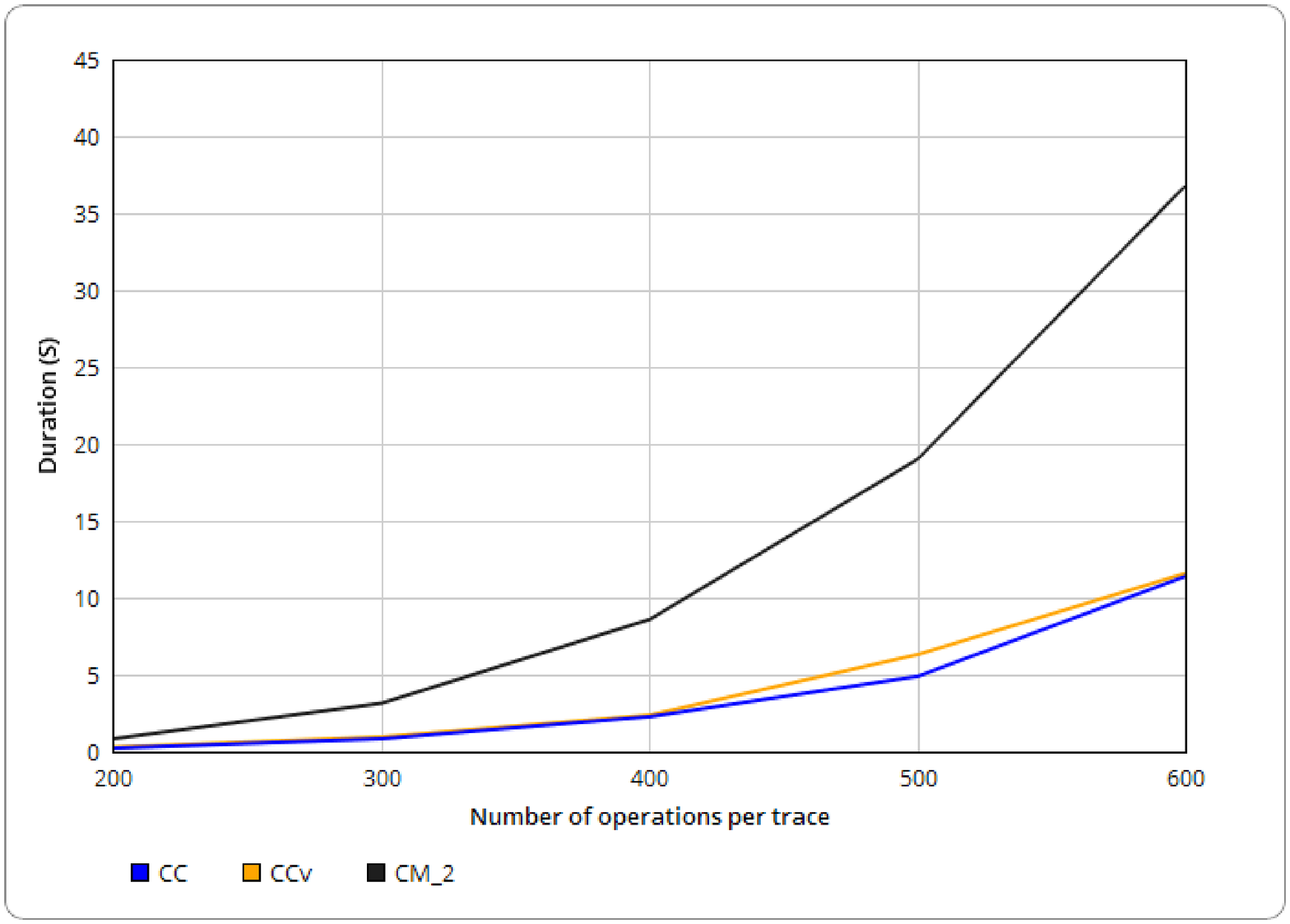} 
\caption{Checking \texttt{CC}, \texttt{CCv} and \texttt{CM_2} while varying the number of operations.}
\label{fig:ccccvnewcm-ops}
\end{minipage}
\hspace{3mm}
\begin{minipage}[t]{0.45\textwidth}
\includegraphics[scale=0.235]{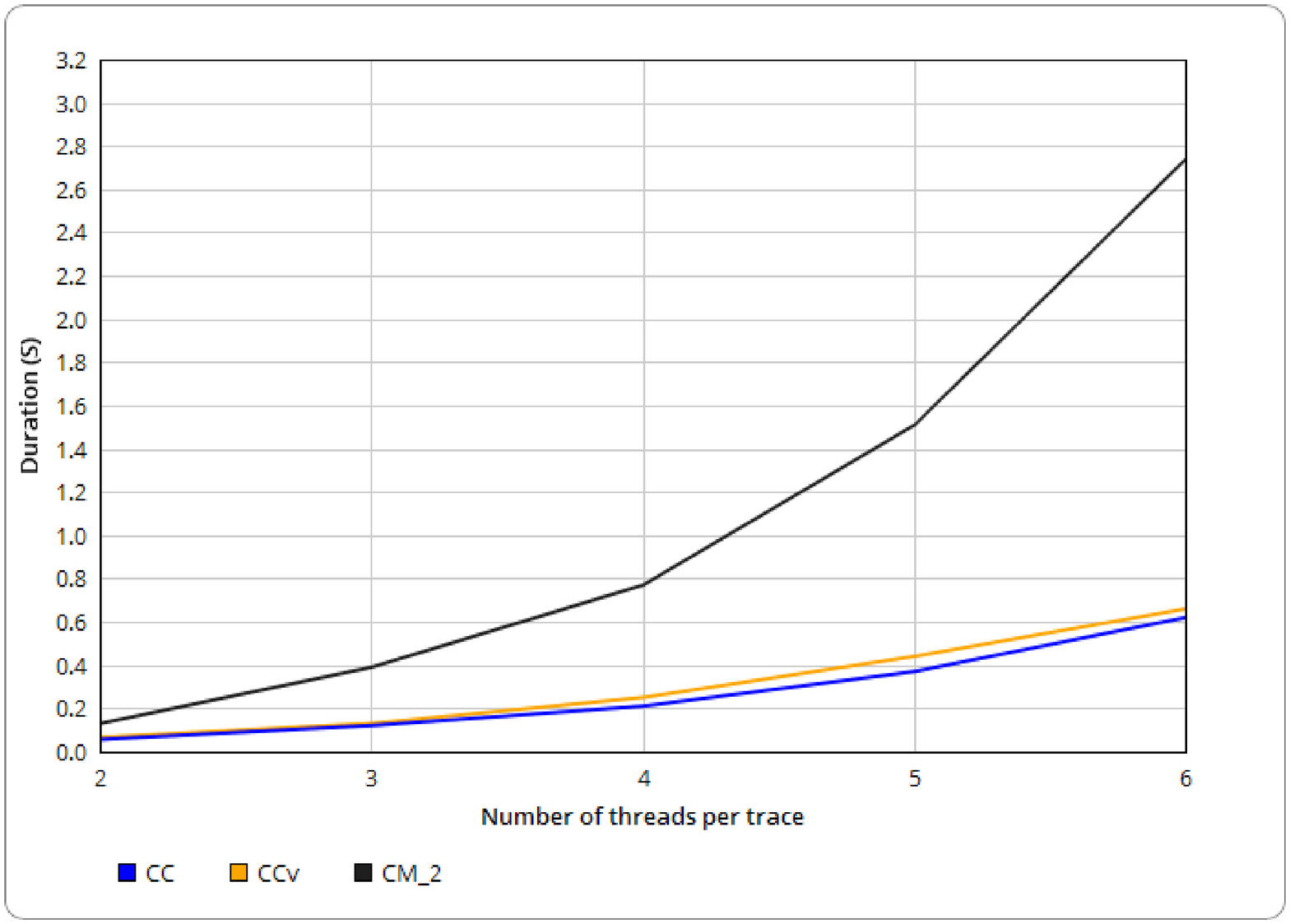} 
\caption{Checking \texttt{CC}, \texttt{CCv} and \texttt{CM_2} while varying the number of processes.}
\label{fig:ccccvnewcm-cpus}
\end{minipage}
\hspace{3mm}
\end{subfigure}
\begin{subfigure}[t]{1\textwidth}
\begin{minipage}[t]{0.45\textwidth}
\includegraphics[scale=0.235]{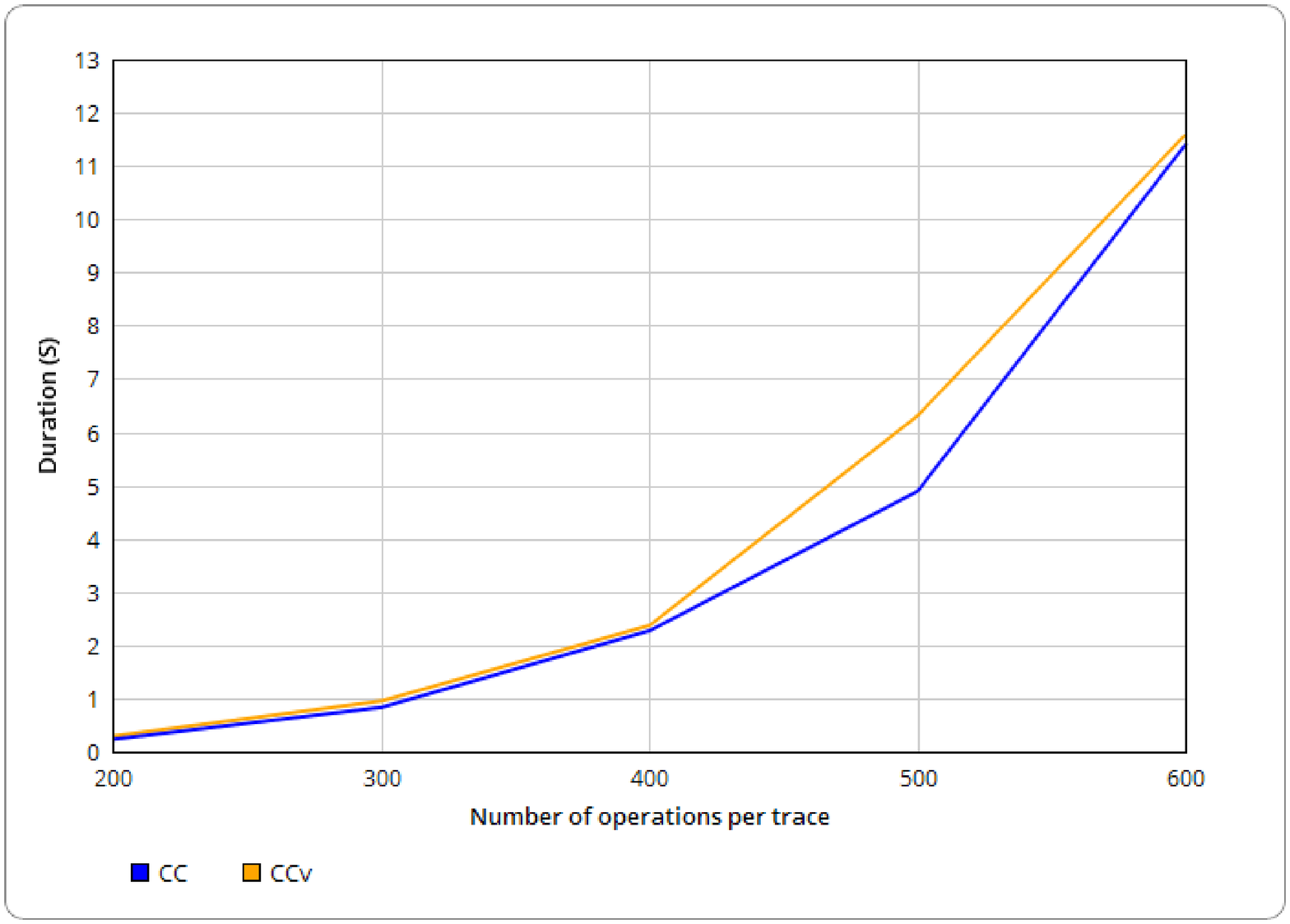} 
\caption{Checking \texttt{CC} and \texttt{CCv} while varying the number of operations.}
\label{fig:ccccv-ops}
\end{minipage}
\hspace{3mm}
\begin{minipage}[t]{0.45\textwidth}
\includegraphics[scale=0.235]{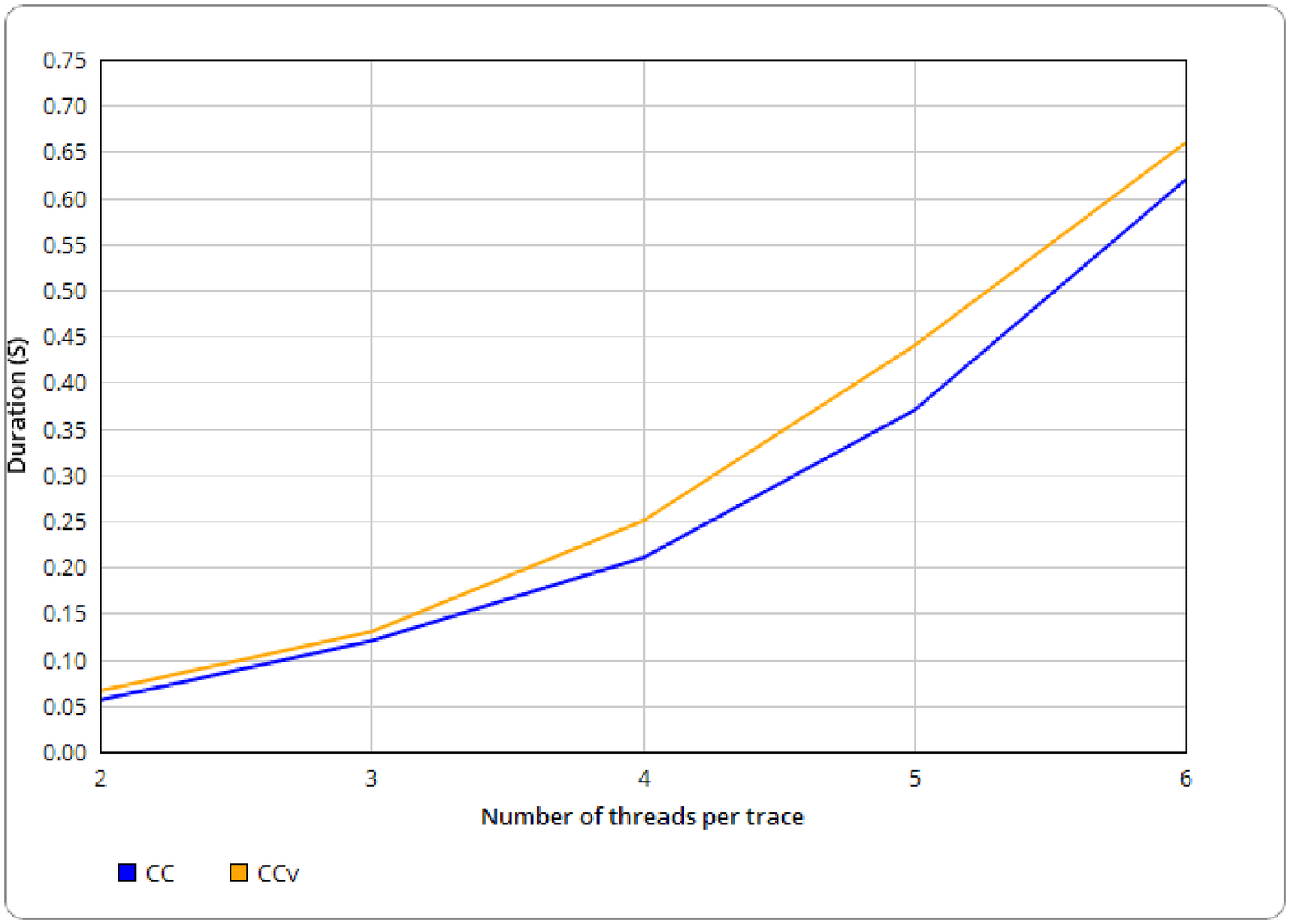} 
\caption{Checking \texttt{CC} and \texttt{CCv} while varying the number of processes.}
\label{fig:ccccv-cpus}
\end{minipage}
\hspace{3mm}
\end{subfigure}
\begin{subfigure}[t]{1\textwidth}
\begin{minipage}[t]{0.45\textwidth}
\includegraphics[scale=0.235]{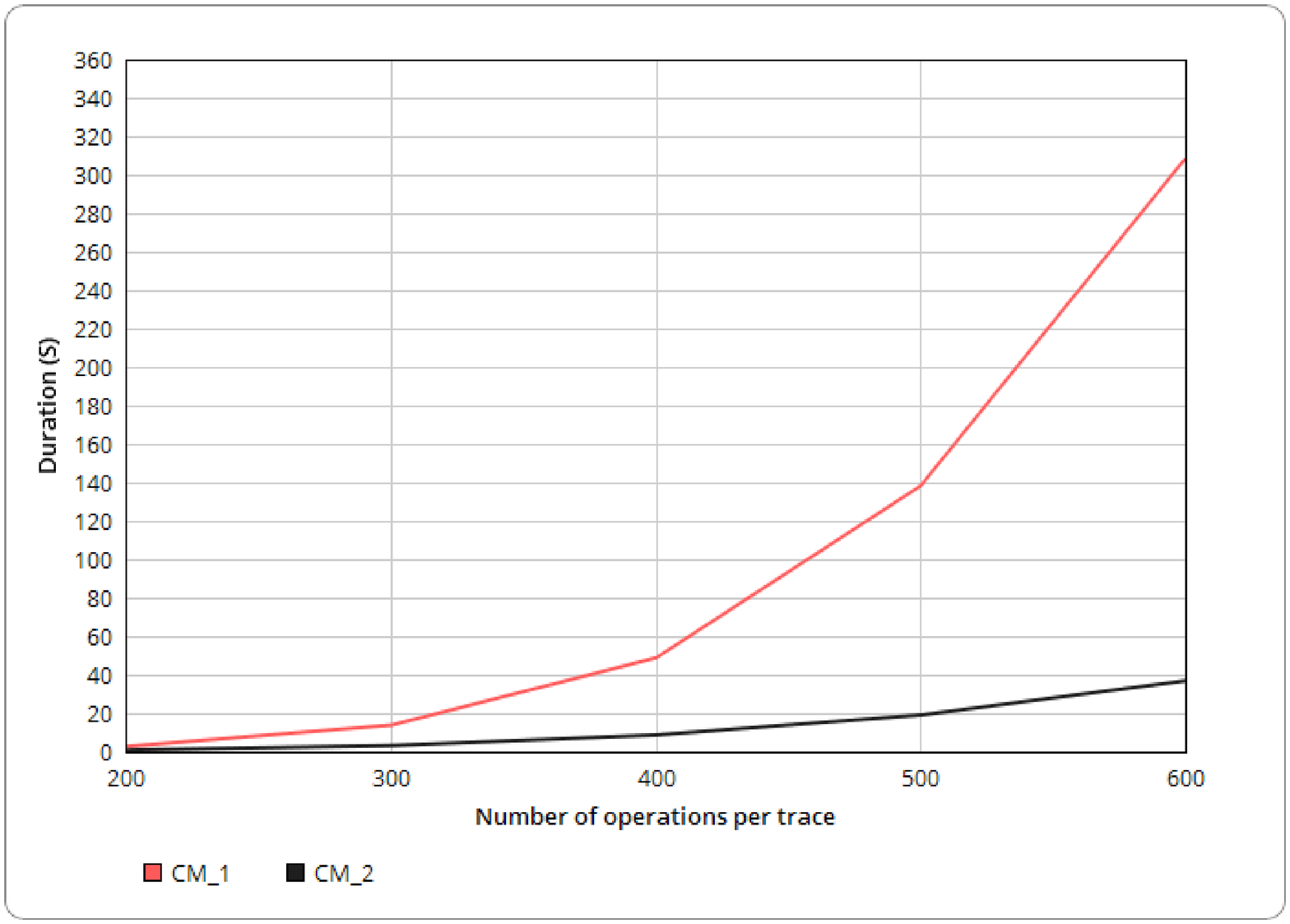} 
\caption{Comparing \texttt{CM_1} and \texttt{CM_2} runtimes while varying the number of operations.}
\label{fig:cmnewcm-ops}
\end{minipage}
\hspace{3mm}
\begin{minipage}[t]{0.45\textwidth}
\includegraphics[scale=0.235]{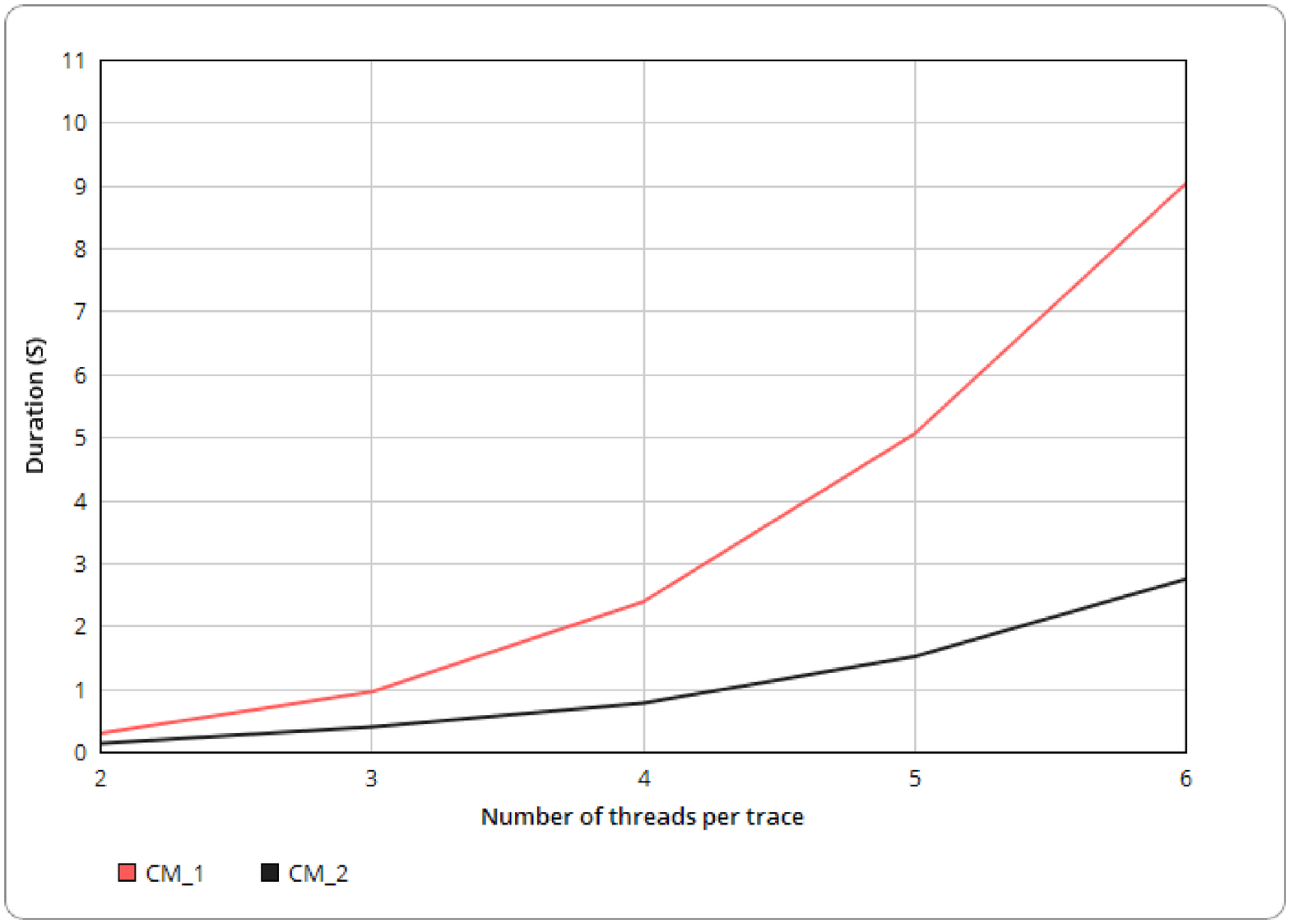} 
\caption{Comparing \texttt{CM_1} and \texttt{CM_2} runtimes while varying the number of processes.}
\label{fig:cmnewcm-cpus}
\end{minipage}
\end{subfigure}
\caption{Checking Causal Consistency for CockreachDB histories.}
\label{fig:all-valid-fig-crdb}
\end{figure*}
We have examined the effect of the number of operations on runtime for a fixed number of processes (4 processes) and the effect of the number of processes. We have tested 200 histories for each configuration and calculated the average runtime.

We have checked  \texttt{CC}, \texttt{CCv} and \texttt{CM}, using its two definitions \texttt{CM_1} and \texttt{CM_2}, for all generated histories. Figure~\ref{fig:all-valid-fig-crdb} shows the results. The graphs ~\ref{fig:ccccvcm-ops}, \ref{fig:ccccvnewcm-ops}, \ref{fig:ccccv-ops} and \ref{fig:cmnewcm-ops} show the runtime while increasing the number of operations from 100 to 600, in augmentations of 100 (with a fixed number of processes, 4 processes). The graphs ~\ref{fig:ccccvcm-cpus} , \ref{fig:ccccvnewcm-cpus}, \ref{fig:ccccv-cpus} and \ref{fig:cmnewcm-cpus} report the runtime when increasing the number of processes from 2 to 6, in augmentations of 1. For each number of processes $x$ we have considered $50x$ operations, so increasing the number of processes increases the number of operations in the history as well.

The graph \ref{fig:ccccvcm-ops} resp., \ref{fig:ccccvcm-cpus} shows a comparaison between \texttt{CC}, \texttt{CCv}, \texttt{CM_1} and \texttt{CM_2} verification runtimes  while varying the number of operations resp., the number of processes.
The graph \ref{fig:ccccvnewcm-ops} resp., \ref{fig:ccccvnewcm-cpus}, presents the running time of \texttt{CM_2} verification compared to \texttt{CC} and \texttt{CCv} verification running time. The graph \ref{fig:ccccv-ops} resp., graph ~\ref{fig:cmnewcm-ops} , shows the evolution of \texttt{CC} and \texttt{CCv} verification resp., \texttt{CM_1} and \texttt{CM_2} verification, runtime while increasing the number of operations. The graph \ref{fig:ccccv-cpus}  resp., graph ~\ref{fig:cmnewcm-cpus}, shows the evolution of \texttt{CC} and \texttt{CCv} verification resp., \texttt{CM_1} and \texttt{CM_2} verification, runtime while increasing the number of processes.

Our approach is more efficient in the case of \texttt{CC} and \texttt{CCv} verification compared to the \texttt{CM_1} case (graphs \ref{fig:ccccvcm-ops} and \ref{fig:ccccvcm-cpus}). The figure \ref{fig:ccccvnewcm-ops} resp., \ref{fig:ccccvnewcm-cpus}, is a zoom on \texttt{CC}, \texttt{CCv} and \texttt{CM_2} of figure \ref{fig:ccccvcm-ops} resp., \ref{fig:ccccvcm-cpus}.  It shows that the \texttt{CM_2} improves the running time but costs more compared to \texttt{CC} and \texttt{CCv} as well. The figure \ref{fig:ccccv-ops} resp., \ref{fig:ccccv-cpus}, is a zoom on \texttt{CC} and \texttt{CCv} of figure \ref{fig:ccccvcm-ops} resp., \ref{fig:ccccvcm-cpus}. It shows that \texttt{CC} and \texttt{CCv} verification are very efficient and terminates in less than 11.6 seconds for all histories we have tested.
As we have noticed above, the results shown in ~\ref{fig:cmnewcm-ops} and \ref{fig:cmnewcm-cpus} show that \texttt{CM_2} has better performance, by factors of 8 times in the case of 600 operations. As expected, all the tested histories were valid w.r.t. all the considered causal consistency models.
\subsection{Case study 2: Galera.}
\begin{figure*}
\footnotesize
\centering
\begin{subfigure}[t]{1\textwidth}
\begin{minipage}[t]{0.45\textwidth}
\includegraphics[scale=0.235]{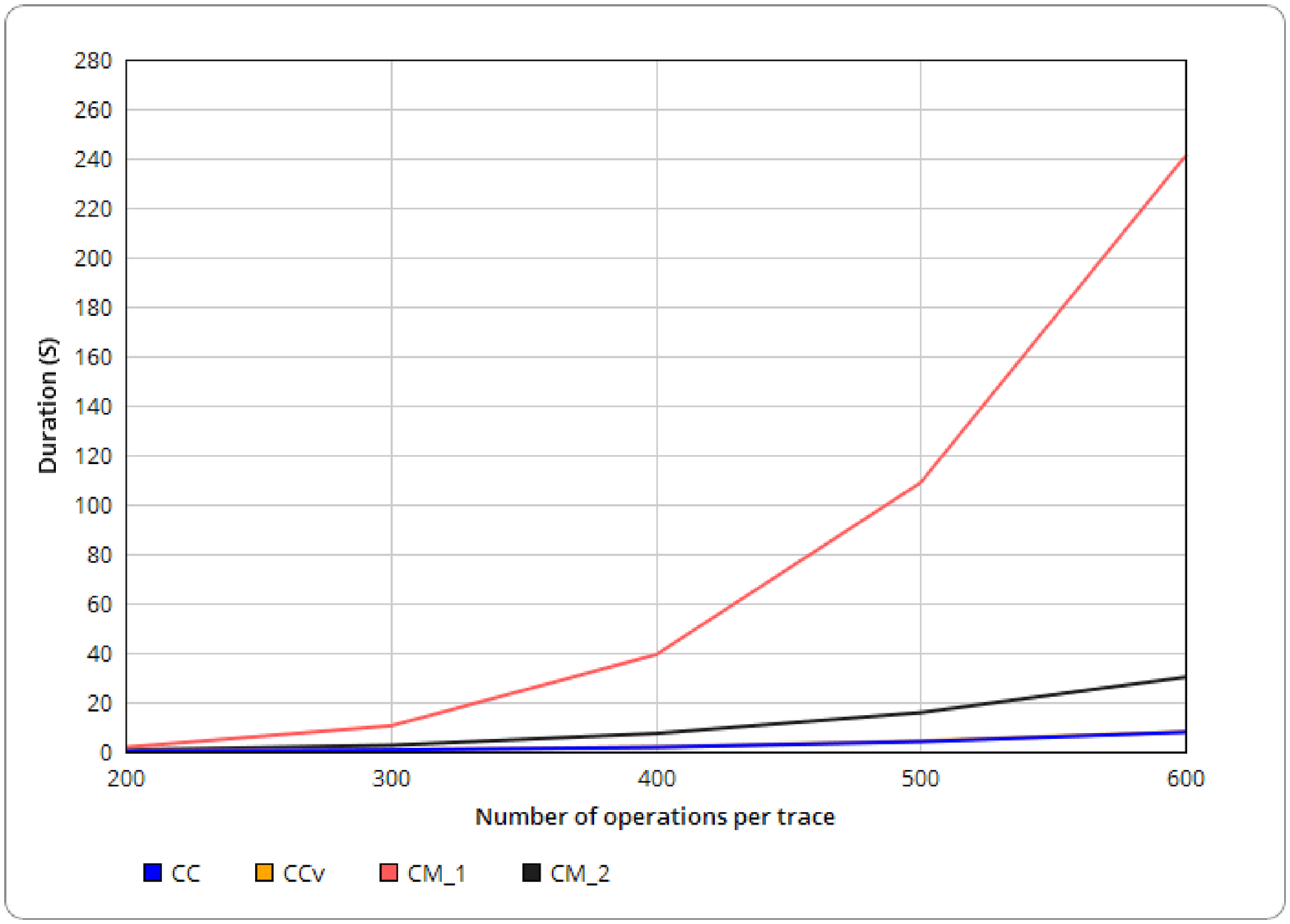} 
\caption{Checking Causal Consistency while varying the number of operations.}
\label{fig:gal-ccccvcm}
\end{minipage}
\hspace{3mm}
\begin{minipage}[t]{0.45\textwidth}
\includegraphics[scale=0.235]{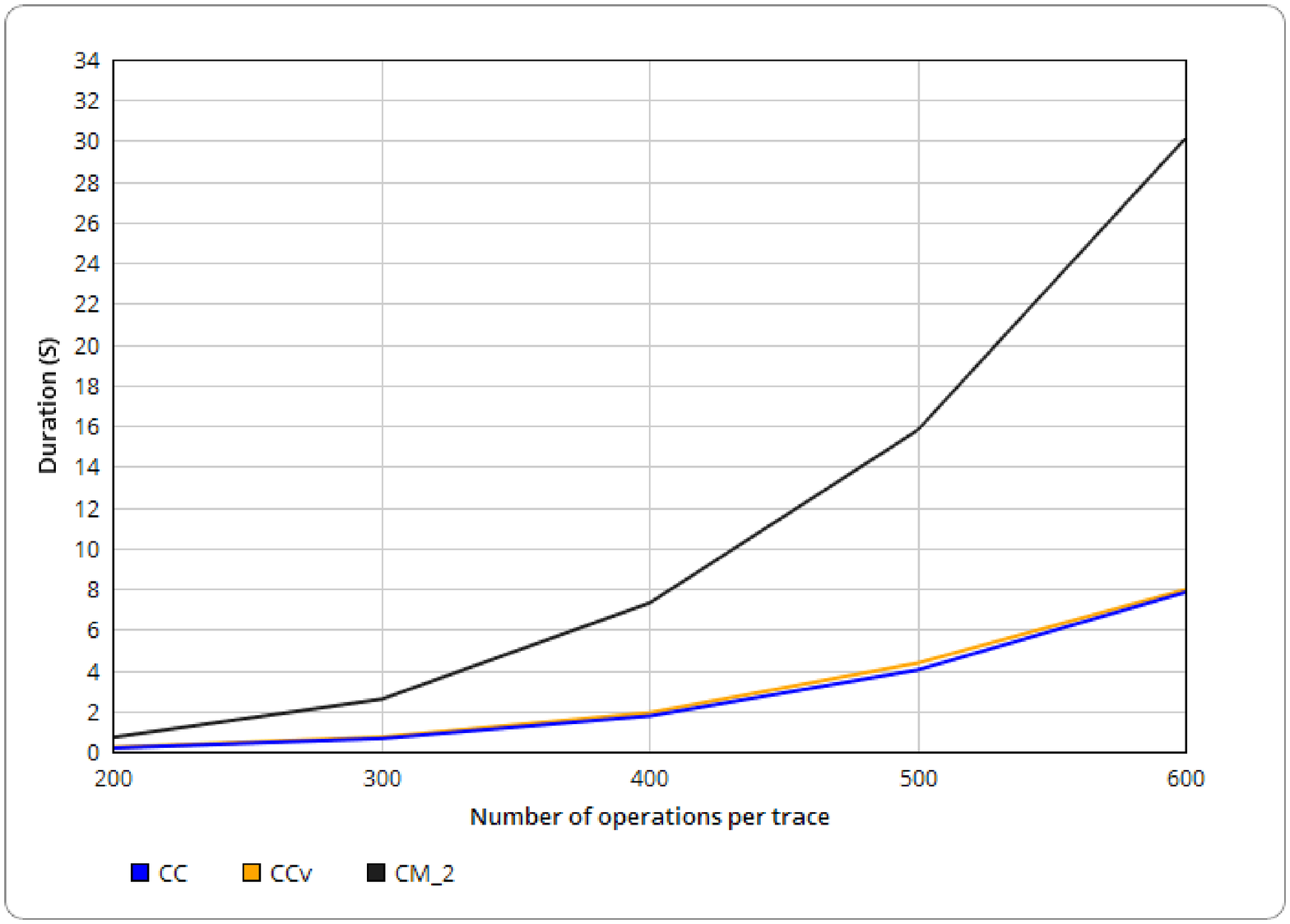} 
\caption{Checking \texttt{CC}, \texttt{CCv} and \texttt{CM_2} while varying the number of operations.}
\label{fig:gal-ccccvnewcm}
\end{minipage}
\hspace{3mm}
\end{subfigure}
\begin{subfigure}[t]{1\textwidth}
\begin{minipage}[t]{0.45\textwidth}
\includegraphics[scale=0.235]{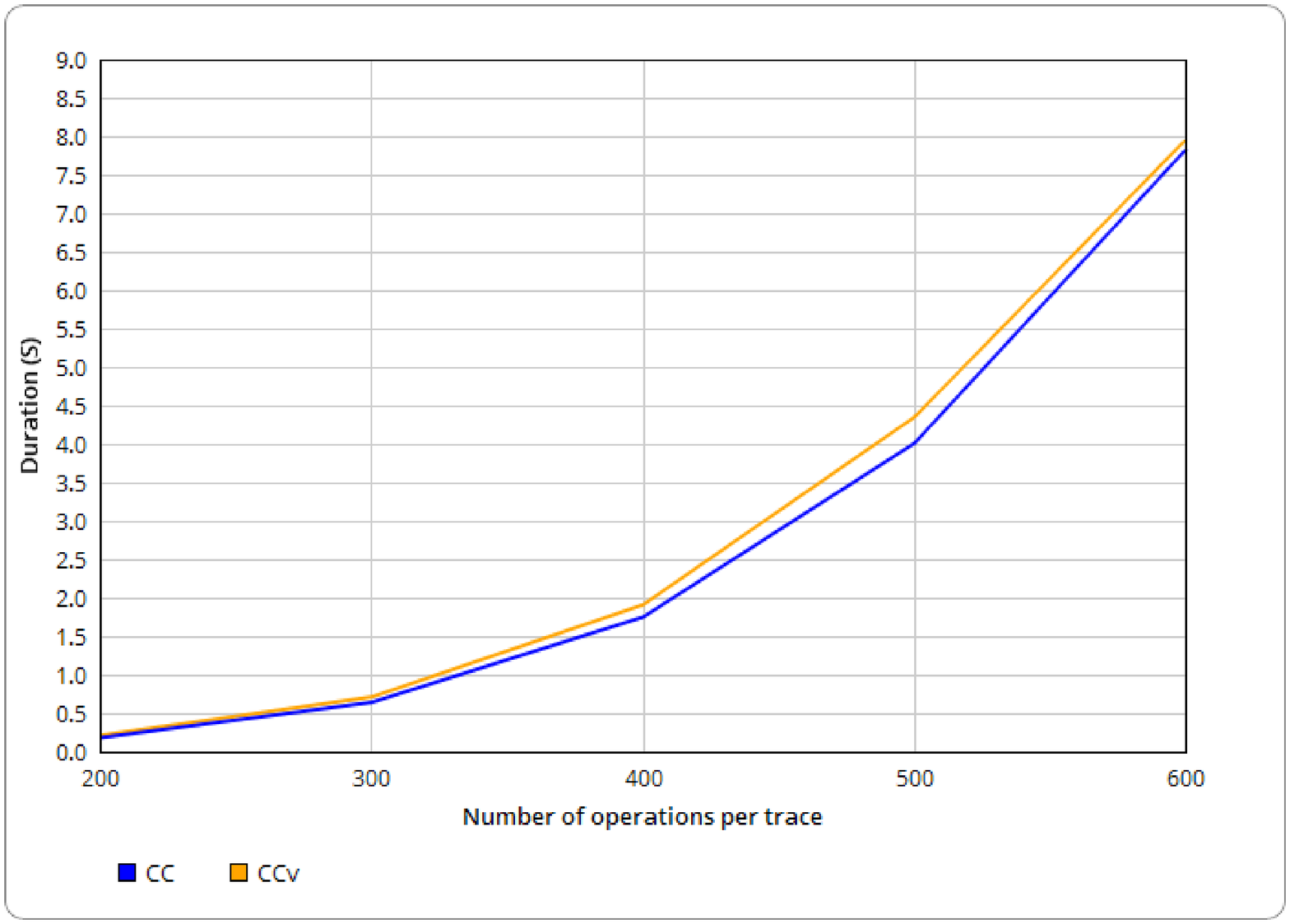} 
\caption{Checking \texttt{CC} and \texttt{CCv} while varying the number of operations.}
\label{fig:gal-ccccv}
\end{minipage}
%\hspace{3mm}
%\begin{minipage}[t]{0.45\textwidth}
%\includegraphics[scale=0.235]{galera-cm-new-cm.eps}
%\caption{Comparing \texttt{CM_1} and \texttt{CM_2} runtimes while varying the number of operations.}
%\label{fig:gal-cmnewcm}
%\end{minipage}
\hspace{3mm}
%\end{subfigure}
%\begin{subfigure}[t]{1\textwidth}
\begin{minipage}[t]{0.45\textwidth}
\includegraphics[scale=0.235]{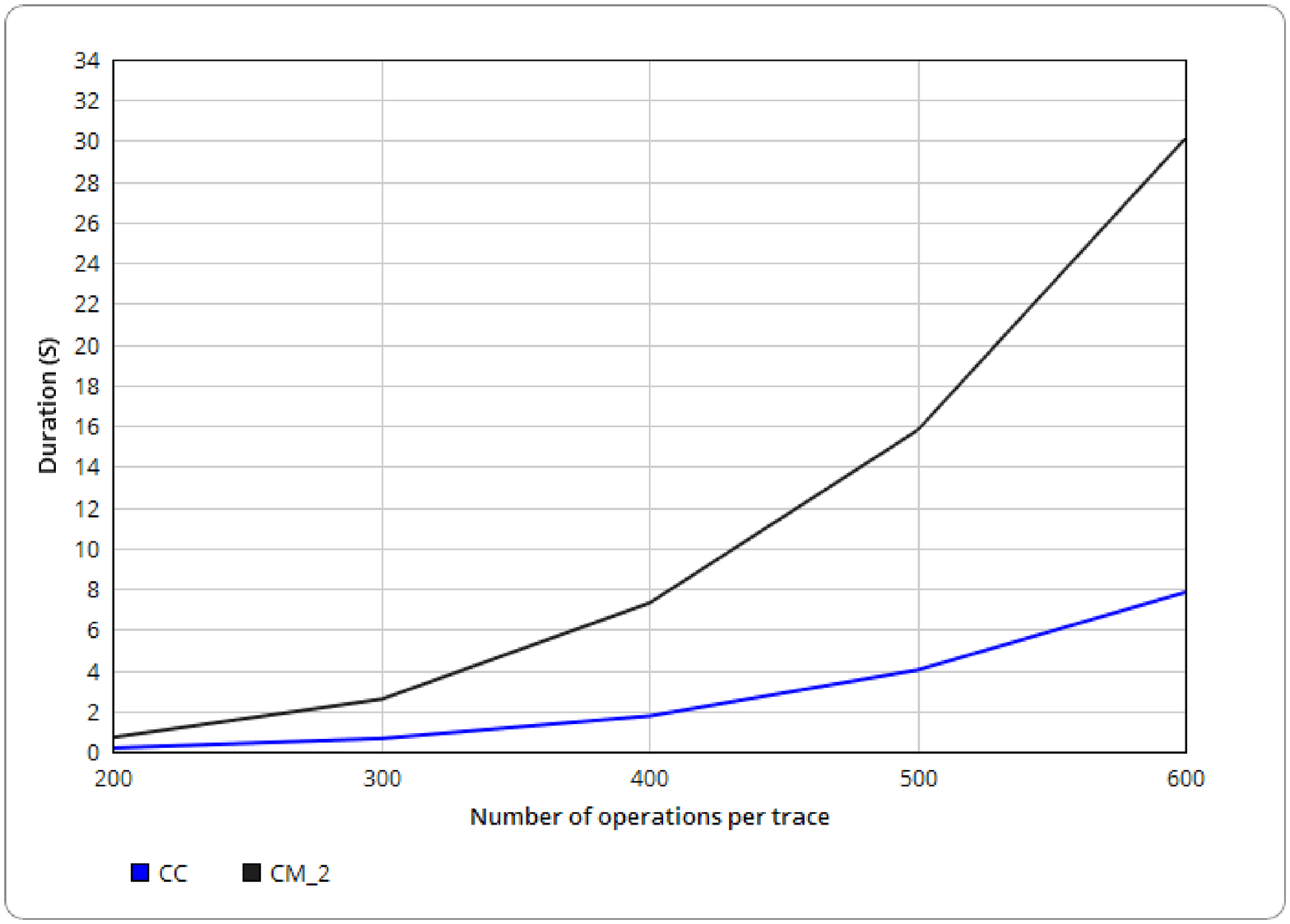} 
\caption{Comparing \texttt{CM_2} and \texttt{CC} violations runtimes while varying the number of operations.}
\label{fig:gal-cc-vs-newcm}
\end{minipage}
\hspace{3mm}
\end{subfigure}
\caption{Checking Causal Consistency for Galera histories.}
\label{fig:all-fig-galeradb}
\end{figure*}
We have also used the cluster called Galera \cite{GALDB} (v3.20). Galera Cluster is a database cluster based on synchronous replication and Oracle’s InnoDB/MySQL. It is expected to implement \textit{Snapshot isolation} when transactions are processed in separated nodes. \\
Similarly to the first case study, we have studied the evolution of runtime while increasing the number of operations from 100 to 600, in augmentations of 100. We have verified 200 histories for each number of operations and compute the runtime average.

The graphs in Figure~\ref{fig:all-fig-galeradb} show the impact of increasing the number of operations on runtime while fixing the number of processes (4 processes). The graph \ref{fig:gal-ccccvcm} shows the comparaison of \texttt{CC}, \texttt{CCv}, \texttt{CM_1} and \texttt{CM_2} verification runtimes. The graph \ref{fig:gal-ccccvnewcm} presents a zoom on graph \ref{fig:gal-ccccvcm} in order to compare \texttt{CM_2} to \texttt{CC} and \texttt{CCv}. The graph \ref{fig:gal-ccccv} reports the evolution of \texttt{CC} and \texttt{CCv} verification runtime. %The graph ~\ref{fig:gal-cmnewcm} shows the evolution of \texttt{CM_1} and \texttt{CM_2} checking runtimes. 
Finally, the graph \ref{fig:gal-cc-vs-newcm} presents a comparison between \texttt{CC} and \texttt{CM_2} running times.

Similarly to the CockroachDB case study, our approach is more efficient in the case of \texttt{CC} and \texttt{CCv} either while increasing the number of operations or processes. The graph %\ref{fig:gal-cmnewcm} 
\ref{fig:gal-ccccvcm} shows that our new definition \texttt{CM_2} outperforms \texttt{CM_1}, but still less efficient compared to \texttt{CC} and \texttt{CCv} (graph \ref{fig:gal-ccccvnewcm}).

Our approach allows capturing violations on the Galera database. We have found that 1.25$\%$ of the tested Galera histories violate causal consistency, that confirms the bugs submitted on Github\cite{ISSUE}. We mention that $73.3\%$ of the detected \texttt{CM} violations are also \texttt{CC} violations. The suggested approach scales well and detects violations on the used version of Galera DB.

The experiments show that our approach is efficient for both verification of valid computations and detection of violations, especially in the case of \texttt{CC} and \texttt{CCv}. The gap between \texttt{CC} (\texttt{CCv}) and \texttt{CM_1} runtimes reported in the graphs~\ref{fig:ccccvcm-ops},~\ref{fig:ccccvcm-cpus} and~\ref{fig:gal-ccccvcm} is due to the fact that in \texttt{CM_1} we compute the $\hb[o]$ relation and check the bad-patterns for each operation. This gap is reduced using the new definition \texttt{CM_2} (graphs \ref{fig:ccccvcm-ops}, \ref{fig:ccccvcm-cpus} and \ref{fig:gal-ccccvcm}) in which  we compute the $\hb[o]$ relation and check the bad-patterns for only the last operation of each thread. 

 %Since \texttt{CM_1} and \texttt{CM_2} cost more compared to \texttt{CC} in terms of runtime (Figures \ref{fig:gal-ccccvcm} and \ref{fig:gal-cc-vs-newcm}) and the most \texttt{CM_1} violations in practice are \texttt{CC} violations ($73.3\%$ in the Galera case), one can start by verifying \texttt{CC} first. 
%!TEX root = main.tex
\section{Related Work}\label{sec-rw}
Several works have considered the problem of checking strong consistency models such as \textit{Linearizability} and \textit{Sequential consistency} (\texttt{SC})~\cite{DBLP:conf/oopsla/ParoshABMTK19,DBLP:journals/sttt/AbdullaHH16,%DBLP:journals/iandc/AlurMP00,
DBLP:conf/pldi/BurckhardtDMT10,DBLP:conf/cav/EirikssonM95,DBLP:journals/pacmpl/EmmiE18,DBLP:conf/pldi/EmmiEH15,%DBLP:journals/jacm/GermanS92,
DBLP:journals/tpds/Qadeer03,DBLP:journals/jpdc/WingG93}. Our recent works \cite{DBLP:conf/cav/ZennouBEE19,DBLP:atva/ZABBEE2020} address the problem of verifying \texttt{SC} and \texttt{TSO} (\textit{Total store ordering}) gradually by using several variants of causal consistency (and other weak consistency models) including the ones we have considered in this work. However, few have addressed the problem of checking weak consistency models.
Emmi and Enea~\cite{DBLP:conf/cav/EmmiE18} propose an algorithm to optimize the  consistency checking based on the notion of minimal-visibility. However, their work relies on some specific relaxations in those criteria, leading to the naive enumeration in the context of strong consistency models such as \texttt{SC} and \texttt{TSO}. Bouajjani et al.~\cite{DBLP:popl/Bouajjani2014} presents a formalization of eventual consistency for replicated objects and reduces the problem of checking eventual consistency to reachability and model checking problems.\\
Bouajjani et al.~\cite{DBLP:conf/popl/BouajjaniEGH17} considers the problem of checking causal consistency. They present the formalization of the different variations of causal consistency(\texttt{CC}, \texttt{CCv} and \texttt{CM}) we use in this work and a complete characterization of the violations of those models. In addition, they show that checking if an execution satisfies one of those models is polynomial time ($\mathcal{O}(n^5)$). However, this work does not propose any implementation. Our work presents an implementation based on a reduction to Datalog queries solving which improves the complexity from $\mathcal{O}(n^5)$ to $\mathcal{O}(n^3)$.
%!TEX root = main.tex
\section{Conclusion}\label{sec-conc}
We have presented a tool for checking automatically that given computations of a system are causally consistent. Our procedure for solving this conformance problem is based on implementing the theoretical approach introduced in~\cite{DBLP:conf/popl/BouajjaniEGH17} where causal consistency violations are characterized in terms of the occurrence of some particular bad-patterns. We build on this work by reducing the problem of detecting the existence of these patterns in computations to the problem of solving Datalog queries. We have applied our algorithm to two real-life case studies. The experimental results show that in the case of \texttt{CC} and \texttt{CCv} our approach is efficient and scalable. In the \texttt{CM} case, the cost grows polynomially but much faster than in the case of \texttt{CC} and \texttt{CCv}. In order to improve the \texttt{CM} checking performance, an optimized definition (CM_2) of the original definition \cite{DBLP:conf/popl/BouajjaniEGH17} has been proposed. Our experimental results show that this new definition reduce considerably the cost of \texttt{CM} verification %and lead to a better conformance checking procedure. 
It reduces the \texttt{CM} verification runtime by more than 7 times (for histories with 600 operations). However, this optimized \texttt{CM} definition still less efficient compared to \texttt{CC} and \texttt{CCv}. Nevertheless, it turned out that interestingly, most of the \texttt{CM} violations (73.3$\%$) that we found are in fact \texttt{CC} violations, and therefore can be caught using a more efficient procedure in which one can start by verifying \texttt{CC} first.

%\bibitem{ref_article1}this
%Author, F.: Article title. Journal \textbf{2}(5), 99--110 (2016)

%Author, F., Author, S.: Title of a proceedings paper. In: Editor,
%F., Editor, S. (eds.) CONFERENCE 2016, LNCS, vol. 9999, pp. 1--13.
%Springer, Heidelberg (2016). \doi{10.10007/1234567890}

%Author, F., Author, S., Author, T.: Book title. 2nd edn. Publisher,

%Author, A.-B.: Contribution title. In: 9th International Proceedings
%on Proceedings, pp. 1--2. Publisher, Location (2010)

%\bibitem{ref_url1
%LNCS Homepage, \url{http://www.springer.com/lncs}. Last accessed 4
%\end{thebibliography}

\bibliographystyle{splncs04}
\bibliography{misc,dblp}

%% Appendix
%\input{appendix}
%\input{summary-of-differences.tex}
\end{document}